\documentclass{nature}
\usepackage[colorlinks,linkcolor={blue},citecolor={blue},urlcolor={red}]{hyperref}
\usepackage{threeparttable}
\usepackage{graphicx}
\usepackage{makecell}
\usepackage{amsmath}
\usepackage{amssymb}    % \leqslant, \lesssim, \mathbb, etc.
\usepackage{url}        % \url
\usepackage[font={small,stretch=1.25}]{caption}
\usepackage{orcidlink}

\spacing{1.4}

\title{Unexpected clustering pattern in dwarf galaxies challenges formation models}
\author{
Ziwen Zhang\orcidlink{0000-0002-9272-5978}$^{1,2}$, 
Yangyao Chen\orcidlink{0000-0002-4597-5798}$^{1,2}$, 
Yu Rong\orcidlink{0000-0002-2204-6558}$^{1,2}$, 
Huiyuan Wang\orcidlink{0000-0002-4911-6990}$^{1,2}$, 
Houjun Mo\orcidlink{0000-0001-5356-2419}$^3$, 
Xiong Luo\orcidlink{0009-0006-0435-9469}$^{1,2}$ 
\&
Hao Li\orcidlink{0000-0002-4326-3543}$^{1,2}$
}

\begin{document}

\maketitle

\begin{affiliations}
 \item Department of Astronomy, University of Science and Technology of China, Hefei, Anhui 230026, China
 \item School of Astronomy and Space Science, University of Science and Technology of China, Hefei 230026, China
 \item Department of Astronomy, University of Massachusetts, Amherst MA 01003-9305, USA
\end{affiliations}

\begin{abstract}
The galaxy correlation function serves as a fundamental tool for studying cosmology, galaxy formation, and the nature of dark matter.
It is well established that more massive, redder and more compact galaxies tend to have stronger clustering in space\cite{Li06a, Zehavi2011}. 
These results can be understood in terms of galaxy 
formation in Cold Dark Matter (CDM) halos of different mass and
assembly history.
Here, we report an unexpectedly strong large-scale clustering for 
isolated, diffuse and blue dwarf galaxies, comparable to that seen for massive galaxy groups but much stronger 
than that expected from their halo mass. 
Our analysis indicates that the strong clustering aligns with the halo assembly bias seen in simulations\cite{Gao2005} 
with the standard $\Lambda$CDM cosmology only if
more diffuse dwarfs formed in low-mass halos of older ages. 
This pattern is not reproduced by 
existing models of galaxy evolution in a $\Lambda$CDM framework 
\cite{Amorisco2016, DiCintio2017, vanDokkum2022Natur}, 
and our finding provides new clues for the search of more viable models.  
Our results can be explained well by assuming self-interacting dark 
matter\cite{Spergel2000PhRvL}, suggesting that such a scenario should be considered seriously.
\end{abstract}

\clearpage

\renewcommand{\figurename}{\bf Fig.}
\renewcommand\thefigure{\arabic{figure}}

\begin{figure}
\includegraphics[scale=0.14]{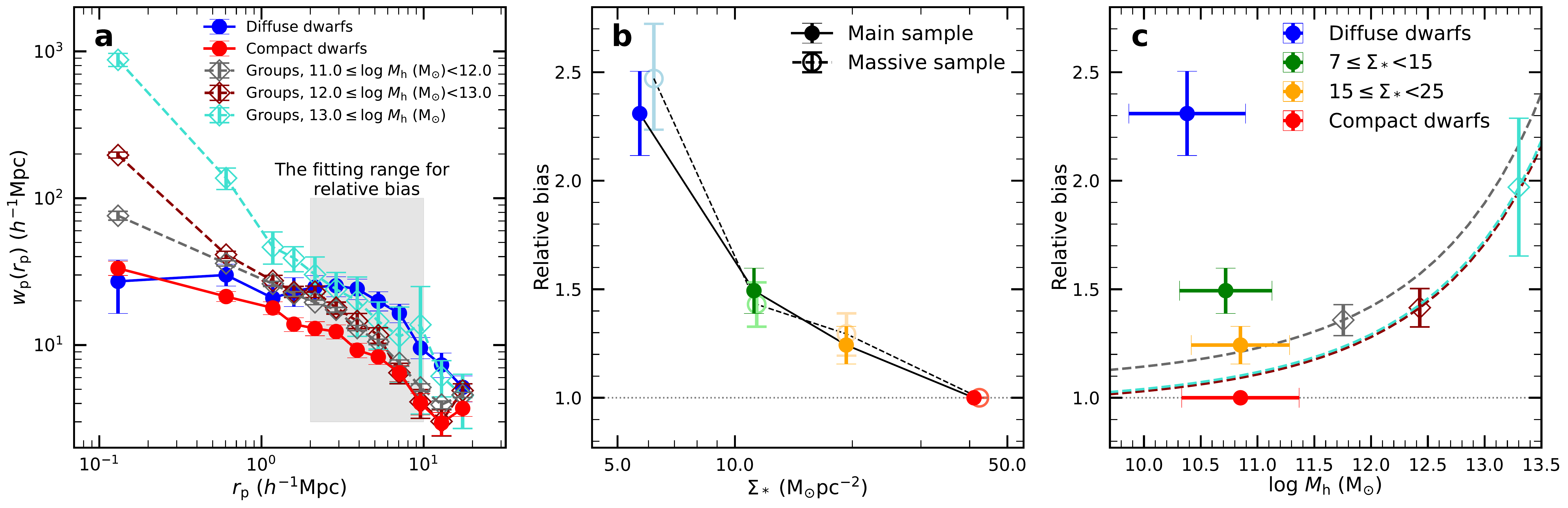}
\centering
\caption{{\bf Projected two-point cross-correlation functions (2PCCFs) 
and relative biases.}
{\bf a}, 2PCCFs ($w_{\rm p}$) as functions of projected separation ($r_{\rm p}$).
Blue and red solid curves are for diffuse and compact SDSS\cite{Abazajian-09} dwarfs,
respectively. Dashed curves are for groups with varying halo masses ($M_{\rm h}$). 
Diffuse dwarfs display the most pronounced large-scale clustering, yet they show 
the least small-scale clustering that is comparable to that of compact dwarfs. 
Shaded region indicates the radial interval used to define the 
large-scale relative bias. 
{\bf b}, Relative bias versus surface mass density ($\Sigma_*$) for dwarfs
(solid, main sample; dashed, massive sample).
A noticeable dependence of bias on $\Sigma_*$ is seen. Note that the relative 
biases for each sample are measured against the compact dwarfs in that sample.
{\bf c}, Relative biases as functions of halo mass for dwarfs and 
galaxy groups\cite{Yang2007}. Dashed curves are the same theoretical 
prediction for the absolute bias\cite{Tinker2010}, scaled to the observed
values of relative biases for groups with different ranges of halo masses.
For comparison, the relative biases for the dwarfs
in the main sample are also shown, with their halo masses obtained by 
HI kinematics. The dependence of bias on halo mass for groups aligns 
with theoretical prediction, but the bias for diffuse dwarfs is 
much higher than that expected from their halo masses. 
The 2PCCFs and the relative biases are computed using the 
$z$-weighting method (see \hyperref[sec_pccf]{Methods}). Markers with error bars 
represent medians with $16^{\rm th}$--$84^{\rm th}$ percentiles
of bootstrap samples for $w_{\rm p}$, and 
of posterior distributions obtained by Markov chain Monte Carlo (MCMC)
fitting for relative biases. 
Markers with error bars for $M_{\rm h}$ of dwarfs
show the medians with dispersions (not uncertainties) 
of the $M_{\rm h}$ distributions.  Markers for $M_{\rm h}$ of groups 
show the medians of the $M_{\rm h}$ distributions.
}\label{fig_bias}
\end{figure}

Our dwarf galaxies are selected from the New York University Value Added Galaxy Catalog sample\cite{Blanton-05a} of the Sloan Digital Sky Survey (SDSS) DR7\cite{Abazajian-09}. We only consider isolated dwarfs, defined as the centrals of galaxy groups\cite{Yang2007}, to avoid complications by satellite galaxies in interpreting our results. We also excluded dwarfs with red color and large S\'ersic index, so that we can focus on ``late-type'' galaxies which have so far been believed to form late and to have weak clustering in space. 
The dwarfs are divided into four samples according to their surface mass density ($\Sigma_*$). We then calculated the projected two-point cross-correlation functions 
(2PCCFs; see \hyperref[sec_pccf]{Methods}), with results shown in Fig.~\ref{fig_bias}a, and derived the relative bias defined 
as the ratio of the 2PCCF of a sample with that of compact (highest-$\Sigma_*$) dwarfs. 
The relative bias as a function of $\Sigma_*$ plotted in Fig.~\ref{fig_bias}b shows clearly  
that the bias increases with decreasing $\Sigma_*$, contrary to common belief. 
For the lowest-$\Sigma_*$ dwarfs (diffuse dwarfs), which are similar to ultra-diffuse galaxies (UDGs)
defined in the literature\cite{vanDokkum2015}, the relative bias is $2.31_{-0.19}^{+0.20}$, indicating 
a dependence on $\Sigma_*$ at about $7\sigma$ level. For the second-lowest $\Sigma_*$ sample, 
the relative bias is $1.49_{-0.11}^{+0.10}$, demonstrating that the decline with $\Sigma_*$ is over 
the entire range of $\Sigma_*$ covered by our sample.

We used various tests to assess the reliability of our findings against effects of sample incompleteness, 
cosmic variance, satellite contamination, and uncertainties in measurements of galaxy properties (see \hyperref[sec_incomp]{Methods}). 
We found that the incompleteness is mainly in $M_*$ and marginally in color and $\Sigma_*$. 
Dividing our sample into two sub-samples with different $M_*$ ranges, we observed no notable difference in the bias-$\Sigma_*$ relation 
between the two. Massive dwarfs with $8.5<\log M_*/{\rm M_\odot}<9$ at $z\leq0.04$ are much more complete than the total population (the main sample), and their results, shown in Fig.~\ref{fig_bias}b for comparison with the main sample, indicate clearly that the incompleteness does not change the 
outcome significantly, as is expected when selection effects are independent of the large-scale 
structure. Dividing the total sample into two sub-volumes either according to sky coverage 
or redshift gives similar results, demonstrating that the cosmic variance does not change our conclusion. 
Stronger clustering would be anticipated if the diffuse dwarf sample were significantly affected by 
satellites in massive groups/clusters of galaxies. This possibility is conclusively negated by examining 
the satellites' contribution, which was found to only increase small-scale correlation 
but have little effects on large scales where the relative bias is measured. 
Finally, uncertainties in galaxy-property measurements are not known to be correlated with 
large-scale structures and thus can only reduce the difference between samples, 
implying that the true correlation between the bias and $\Sigma_*$ is even stronger 
than what is estimated from the data. All these demonstrate that our results are robust against 
observational effects.

Massive halos are known to be clustered more strongly than low-mass halos on average\cite{Mo1996}. 
It is thus interesting to check whether the difference in clustering between the diffuse and compact dwarfs 
is caused by a difference in halo mass.
Here, we present halo mass measurements using two different methods (see \hyperref[sec_shmr]{Methods}). 
The first is based on the rotational velocity traced by HI-emission lines\cite{Hu2023}.
The median halo masses for the diffuse and compact dwarfs with HI detections are $10^{10.38} \rm M_{\odot}$ 
and $10^{10.85} \rm M_{\odot}$, respectively. The second is based on the assumption that different subsets of 
the dwarf population  obey the same stellar mass-halo mass relation (SHMR)\cite{Kravtsov2018}, 
and the median halo masses for diffuse and compact dwarfs obtained in this way are $10^{10.83} \rm M_{\odot}$ 
and $10^{11.01} \rm M_{\odot}$, respectively. The two methods give a consistent result
that both diffuse and compact dwarfs have comparable halo masses. 
The halo bias model\cite{Tinker2010} predicts a bias ratio of $0.94$ and $0.99$
between the diffuse and compact dwarfs using halo masses given by 
the HI kinematics and the SHMR, respectively. 
Even though the uncertainty in the halo mass is large, the uncertainty in the predicted bias ratio is very small (less than or equal to $0.02$),
because the average bias depends very weakly on halo mass in the 
low-mass end\cite{Mo1996, Tinker2010}. Indeed, even we use the upper bound in 
the scatter of the halo masses ($M_{\rm h}=10^{11.5}\, {\rm M_\odot}$) for diffuse dwarfs 
and the lower bound  ($10^{10.0}\, {\rm M_\odot}$) for compact dwarfs, the predicted relative bias 
is only about $1.14$, much lower than the observed value $\sim 2.31$, indicating that the 
difference in clustering between the diffuse and compact samples 
cannot be explained by the difference in their halo masses (see Fig.~\ref{fig_bias}c).  

The clustering of galaxy groups aligns with the halo bias model and simulation predictions\cite{WangE2018b}, 
making it a reliable reference for the absolute clustering strength of dwarf galaxies. 
Fig.~\ref{fig_bias}a shows that, on scales $r_{\rm p}\sim0.1\, h^{-1}{\rm {Mpc}}$, the 
correlation functions for diffuse and compact dwarfs are similar and considerably lower than 
that for groups with $M_{\rm h}\sim10^{11.5}\, {\rm M_\odot}$. Since the small-scale clustering is sensitive to halo mass, 
the result suggests that both diffuse and compact dwarfs inhabit halos with masses below 
$10^{11.5}\, {\rm M_\odot}$, consistent with the halo mass estimates shown above.  However, diffuse dwarfs exhibit much 
stronger clustering on large scales than these groups, with a correlation amplitude    
comparable to that of massive groups with $M_{\rm h}\sim10^{13.5}\, {\rm M_\odot}$ (Fig.~\ref{fig_bias}c). 
These results clearly contradict the conventional expectation that low-mass, blue, and diffuse 
galaxies have weaker clustering than their massive, red, and compact counterparts\cite{Li06a,Zehavi2011}.

\begin{figure}
\includegraphics[width=\textwidth]{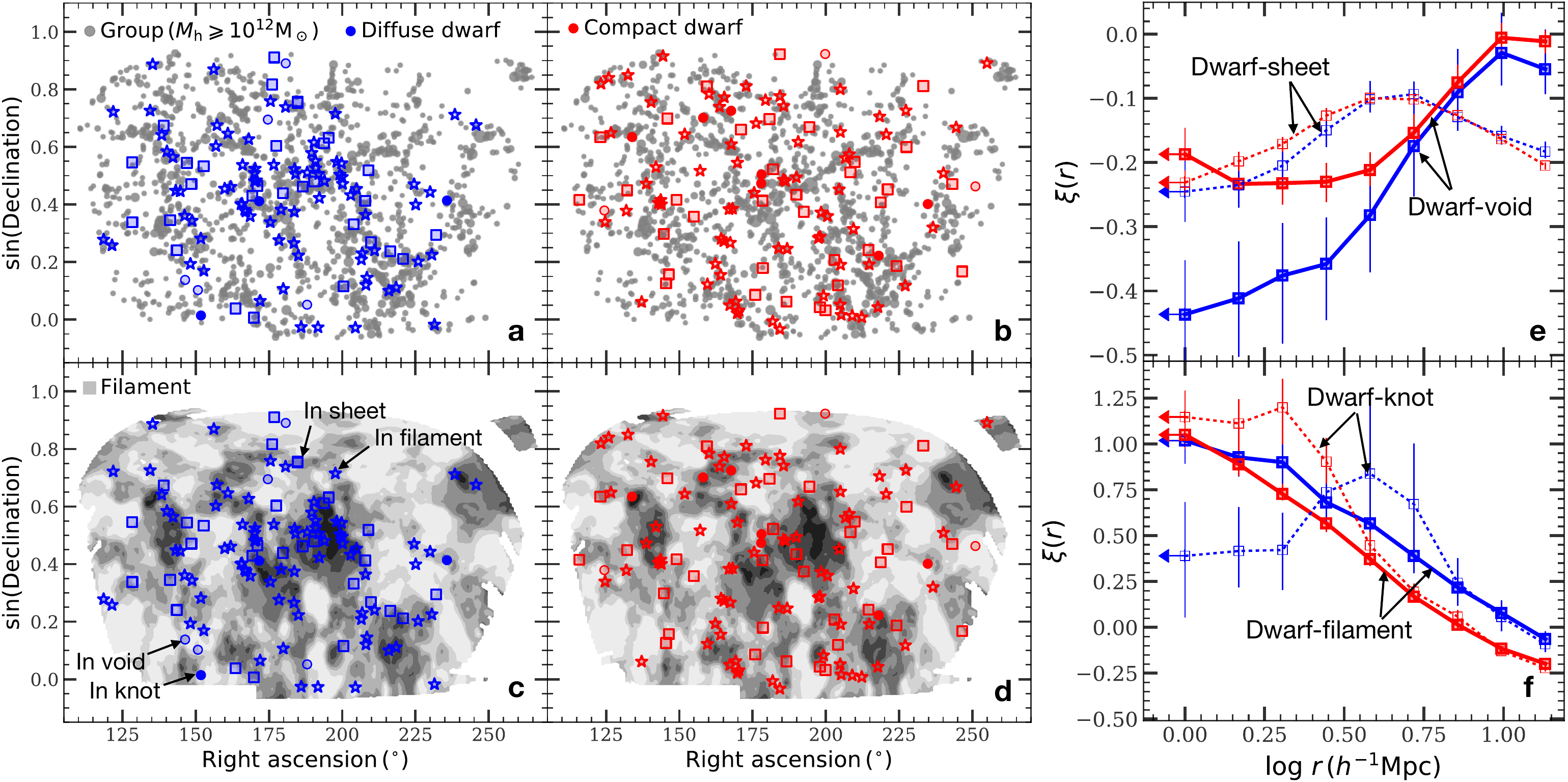}
\centering
    \caption{
    {\bf Correlation between dwarf galaxies and cosmic web.} 
    {\bf a--d}, Spatial distribution of dwarf galaxies, galaxy groups, 
    and filaments of the cosmic web (see \hyperref[sec_cosmic_web]{Methods}). 
    Each blue (red) marker represents a diffuse (compact) dwarf, with 
    different marker shape indicating different type of cosmic web in which it resides.
    Each grey dot in {\bf a} and {\bf b} represent a galaxy group with 
    $M_{\rm h} \geqslant 10^{12}\, {\rm M_\odot}$, with marker size proportional to 
    halo virial radius.
    Grey shades in {\bf c} and {\bf d} show the fraction of field points classified 
    as filament along line of sight, darker for higher fraction.
    Only dwarfs, groups and field points with corrected redshift
    $0.02 < z_{\rm cor} < 0.03$
    are included. Compact dwarfs are down-sampled without replacement to 
    match the number of diffuse dwarfs.
    {\bf e, f}, Real-space 2PCCFs ($\xi$) between dwarfs 
    (blue and red for the diffuse and compact, respectively) and 
    cosmic web of different types (void and sheet in {\bf e}; filament 
    and knot in {\bf f}). Dwarfs are taken from the main sample and are 
    $z$-weighted, while cosmic web points are weighted by 
    their matter density.
    Markers with error bars show medians with $16^{\rm th}$--$84^{\rm th}$ 
    percentiles estimated from bootstrap samples. Leftmost markers 
    (indicated by left arrows) are obtained by combining all pairs
    below $1\, h^{-1}{\rm {Mpc}}$, the smoothing scale of the reconstructed field.
    The strong large-scale correlation with filaments/knots and small-scale 
    anti-correlation with voids of diffuse dwarfs suggest that
    they preferentially reside within/around large cosmic structures.
    }
\label{fig_cosmicweb}
\end{figure}

Fig.~\ref{fig_cosmicweb}a--d depict spatial distributions of diffuse and compact dwarf galaxies
on top of the distribution of galaxy groups\cite{Yang2007} and on filamentary structures\cite{wangELUCIDEXPLORINGLOCAL2016}. 
It appears that diffuse dwarfs tend to be associated with prominent filamentary structures, whereas 
compact dwarfs have a more diffused distribution. To quantify this, we used the reconstructed 
mass density field from the ELUCID project\cite{wangELUCIDEXPLORINGLOCAL2016} to classify the cosmic web 
into filament, sheet, void and knot components.  
Approximately $50\%$ of the dwarfs are found in filaments and $30\%$ in sheets, with diffuse ones 
showing a stronger tendency to reside in filaments than compact ones. We calculated the 2PCCFs between diffuse/compact dwarfs and different 
components of the cosmic web (Fig.~\ref{fig_cosmicweb}e and f). 
Compared to their compact counterparts, diffuse dwarfs show a much weaker correlation with voids, 
but exhibit a stronger association with filaments and knots on large scales. 
This suggests that diffuse dwarfs are more likely to be found within and around large cosmic structures 
than compact dwarfs. However, on small scales, diffuse dwarfs have a weaker correlation with knots 
than compact ones, likely because star-forming gas in diffuse dwarfs is more susceptible to 
stripping by high-density environments than that in compact dwarfs.

\begin{figure} \centering
    \includegraphics[width=0.875\textwidth]{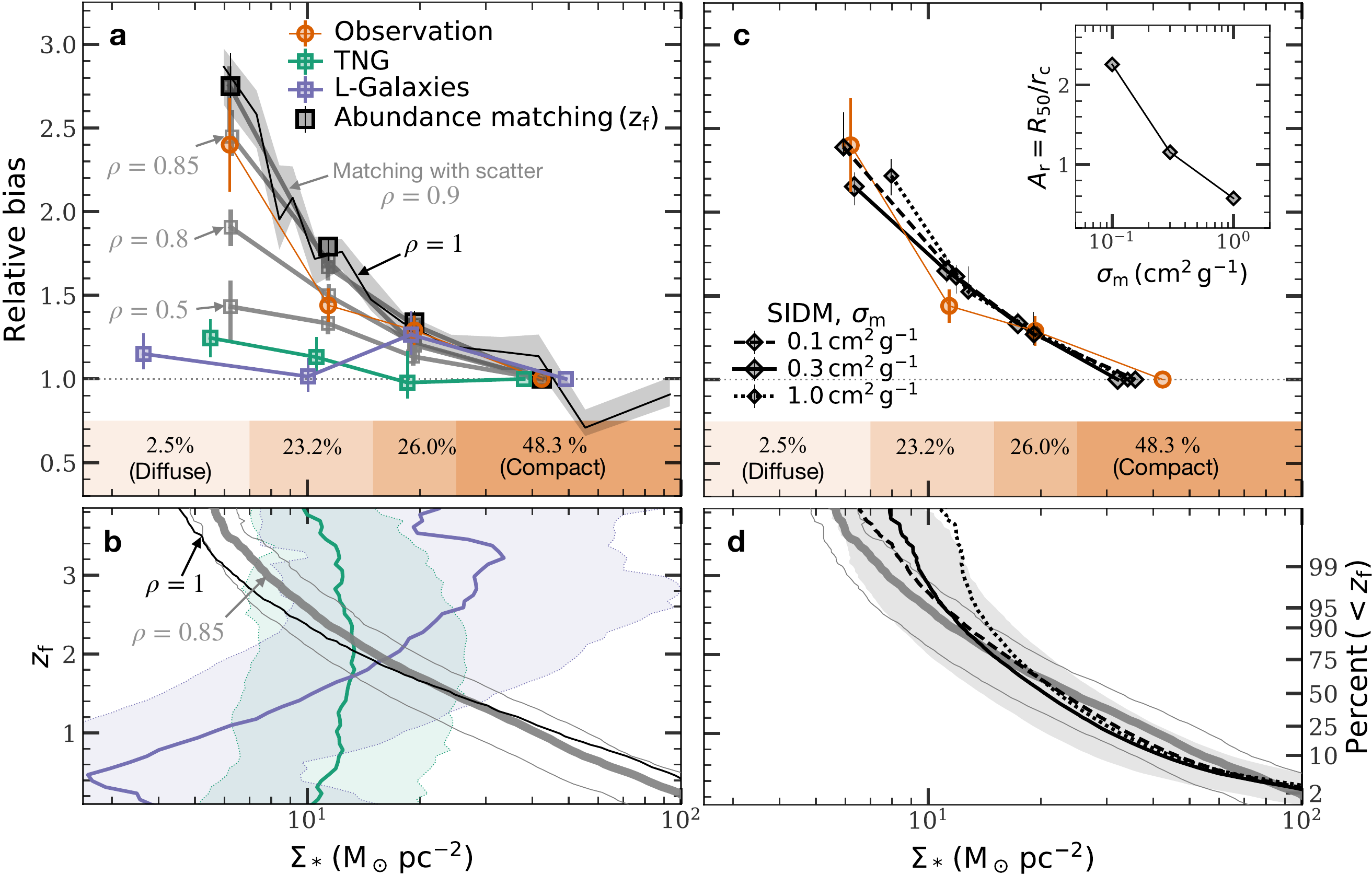}
    \caption{
    {\bf Relative bias as a function of $\Sigma_*$ from galaxy formation models.}
    {\bf a}, bias-$\Sigma_*$ relations from observation (orange)
    and models: TNG\cite{pillepichFirstResultsIllustrisTNG2018} (green), 
    L-Galaxies\cite{ayromlouComparingGalaxyFormation2021} (purple), 
    and our abundance matching (see \hyperref[sec_bias_in_sim]{Methods}) that 
    links $z_{\rm f}$ of halos in the constrained simulation of 
    ELUCID\cite{wangELUCIDEXPLORINGLOCAL2016} to $\Sigma_*$ of observed dwarfs.
    Random scatter in abundance matching is controlled by the correlation coefficient 
    $\rho$ between $\Sigma_*$ and $z_{\rm f}$: 
    $\rho = 1$ for zero scatter (black) and $\rho < 1$ for non-zero scatter (grey).
    Black curve with shades shows a fine-binning result, while black and grey 
    markers are binned the same way as the observation by $\Sigma_*$ 
    (indicated by orange regions, with the percentages of samples labeled).
    {\bf b}, $z_{\rm f}$-$\Sigma_*$ relations 
    implied by the models: our abundance matching with
    $\rho=1$ (black) and $\rho = 0.85$ (grey curve with two bounds,
    also shown in {\bf d}), TNG and L-Galaxies. 
    Panels 
    {\bf c, d} are for the SIDM model assuming different $\sigma_{\rm m}$.
    In {\bf d}, right axis shows the cumulative percentages of $z_{\rm f}$ of 
    ELUCID halos; shades are for the $\sigma_{\rm m}=0.3\,{\rm cm^2g^{-1}}$ case;
    inset panel shows the ratio between galaxy size ($R_{50}$) and 
    SIDM core size ($r_{\rm c}$) required to match the observation, 
    as a function of $\sigma_{\rm m}$.
    The massive sample (see Extended Data Table~\ref{tab_gal}) is used for observation
    and abundance matching.
    Massive ($M_{\rm *} = 10^{8.5}$--$10^{9}\, {\rm M_\odot}$) star-forming central
    dwarfs are used for TNG and L-Galaxies. 
    ELUCID halos with $M_{\rm h} = 10^{10.5}-10^{11}\, {\rm M_\odot}$, 
    within $0.01 \le z \le 0.04$, and without backsplash, are used in 
    abundance matching and SIDM.
    Markers with error bars/shades in {\bf a} and {\bf c} show 
    medians with $16^{\rm th}$--$84^{\rm th}$ percentiles.
    Curves with shades/bounds in {\bf b} and {\bf d} show 
    medians with $16^{\rm th}$--$84^{\rm th}$ percentiles of $\Sigma_*$ at given $z_{\rm f}$.
    Our findings suggest that halo assembly ($z_{\rm f}$) bias
    is sufficient to explain the observed bias-$\Sigma_*$ relation of isolated 
    dwarfs, provided that $\Sigma_*$ has a tight anti-correlation with $z_{\rm f}$.
    }
    \label{fig_b_model}
\end{figure}

For a given mass, the large-scale clustering of halos can also depend on their intrinsic properties, 
a phenomenon referred to as the assembly bias\cite{Gao2005,Wechsler2006, Jing2007, Bett2007}.
The strong dependence of the relative bias on $\Sigma_*$ aligns with such bias provided
that $\Sigma_*$ is correlated with some intrinsic properties of halos.  
Dwarf galaxies are ideal for studying the assembly bias because the dependence of clustering on the halo mass is very weak at the low-mass end. 
We considered two halo properties for which the assembly bias has been investigated extensively:
the spin and the formation redshift $z_{\rm f}$, with the
latter found to be closely correlated with the halo concentration\cite{Gao2004}. 
We found that, for $M_{\rm h}\sim10^{11}\, {\rm M_\odot}$, the dependence of the bias on halo spin
is too weak\cite{Sato-Polito2019} to explain the range of the relative bias shown in
Fig.~\ref{fig_bias}, while the dependence on $z_{\rm f}$ may be sufficient to cover the range
\cite{Wang2009} (see \hyperref[sec_bias_in_sim]{Methods}). 
To quantify this, we first applied the abundance-matching technique to establish 
a connection between $\Sigma_*$ and $z_{\rm f}$ using the massive dwarf 
sample (Fig.~\ref{fig_b_model}b), 
and then assigned a $\Sigma_*$ value to each simulated halo according to its $z_{\rm f}$ and  
the $\Sigma_*$-$z_{\rm f}$ relation. Fig.~\ref{fig_b_model}a shows the relative bias as a
function of $\Sigma_*$ obtained from halos, taken from the constrained simulation of ELUCID\cite{WangH2014,wangELUCIDEXPLORINGLOCAL2016}, in the same volume as the observational sample to minimize
cosmic variances. The observed bias-$\Sigma_*$ relation is well reproduced provided that $\Sigma_*$ is tightly related to $z_{\rm f}$, with a correlation coefficient $\rho >0.8$. 
The question is whether such a relation between $\Sigma_*$ and $z_{\rm f}$ is expected in the current paradigm of galaxy formation.  

In the current cold dark matter (CDM) paradigm, several mechanisms have been proposed for the formation 
of diffuse dwarfs. Environmental processes such as tidal heating, galaxy interaction and ram pressure stripping are found to be able to make dwarf galaxies more  
diffuse\cite{vanDokkum2016, Safarzadeh2017, Jiang2019, liaoUltradiffuseGalaxiesAuriga2019,vanDokkum2022Natur}. 
However, these mechanisms are effective mainly in group and cluster environments, although some simulations suggest that filamentary environments might also strip gas from 
dwarf galaxies\cite{Benitez-Llambay2013}. Such mechanisms are expected to remove gas
from dwarf galaxies and quench star formation in them, producing red and gas-poor dwarfs observed in clusters and groups of galaxies. They are not expected to be efficient for
the formation of the diffuse dwarfs concerned here, because those dwarfs reside in low-mass halos, have blue colors, and possess extended HI disks (see \hyperref[sec_HImass]{Methods}).
It has also been proposed that diffuse dwarfs may be produced in halos of high 
spin\cite{Amorisco2016,RongY2017, liaoUltradiffuseGalaxiesAuriga2019,Benavides2023} according to the disk formation model\cite{Mo1998}. However, this scenario cannot  
explain the strong large-scale clustering of diffuse dwarfs. Alternatively, multiple episodes of supernova
feedback may trigger oscillations in the gravitational potential, which lead to
expansion in the inner parts of halos and the formation of blue diffuse
dwarfs\cite{DiCintio2017,Chan2018}. Such a process might explain the 
observational result if its effect is more significant in older halos. Unfortunately, 
existing simulations suggest that the effect is independent of halo age and 
concentration\cite{DiCintio2017} (see \hyperref[sec_dis_dd_formation]{Methods}). 
The same conclusion can be reached by comparing the observational results 
with the predictions of L-Galaxies\cite{guoDwarfSpheroidalsCD2011,ayromlouComparingGalaxyFormation2021}, a semi-analytic model 
of galaxy formation, and IllustrisTNG\cite{pillepichFirstResultsIllustrisTNG2018} (hereafter TNG), 
a hydro cosmological simulation of galaxy formation.   
These two models do not predict any significant dependence of the bias on $\Sigma_*$
(Fig.~\ref{fig_b_model}a).  
Furthermore, the $z_{\rm f}$-$\Sigma_*$ relation predicted by the two models is 
either very weak or opposite to that needed to explain the bias-$\Sigma_*$ relation   
(Fig.~\ref{fig_b_model}b). 

It is interesting to note that the supernova-driven expansion was proposed as a  
potential solution to the ``small-scale crises'' of the CDM model, such as the 
cusp-core problem and the too-big-to-fail problem\cite{Spergel2000PhRvL,Bullock2017ARAA}. 
However, such a scenario has yet to be extended so as to produce a relation between the expansion and the 
halo assembly in order to explain the observed bias-$\Sigma_*$ relation, and further research 
is needed to assess the feasibility.

Beyond CDM, self-interacting dark matter (SIDM) model has also been proposed as
a promising solution to the small-scale problems\cite{Spergel2000PhRvL, Tulin2018PhR, Kaplinghat2020,yangSelfInteractingDarkMatter2020,zhangSelfinteractingDarkMatter2024}. 
SIDM halos are expected to have
the same formation histories and large-scale clustering as their CDM counterparts, 
so that the assembly bias is also expected to be the same, and
have significantly reduced central densities due to subsequent collisions of
dark matter particles\cite{Rocha2013}. 
Since the probability of collision between dark matter particles increases with
density and halo age, older halos are expected to possess larger cores and lower central
densities\cite{Jiang2023}. Thus, if dwarf galaxies with lower $\Sigma_*$ are  
associated with SIDM halos with larger cores (lower central densities),
as is consistent with the observation that halos of diffuse dwarfs 
usually have low central densities or large cores\cite{Kong2022ApJ,ManceraPina2024},
an anti-correlation between $\Sigma_*$ and $z_{\rm f}$, 
as well as between $\Sigma_*$ and the relative bias are expected, as shown in Fig.~\ref{fig_b_model}a and b.
Thus, the SIDM model combined with the assembly bias provides a plausible explanation for 
the observed bias-$\Sigma_*$ relation. 

\begin{figure*} \centering
    \includegraphics[width=0.825\textwidth]{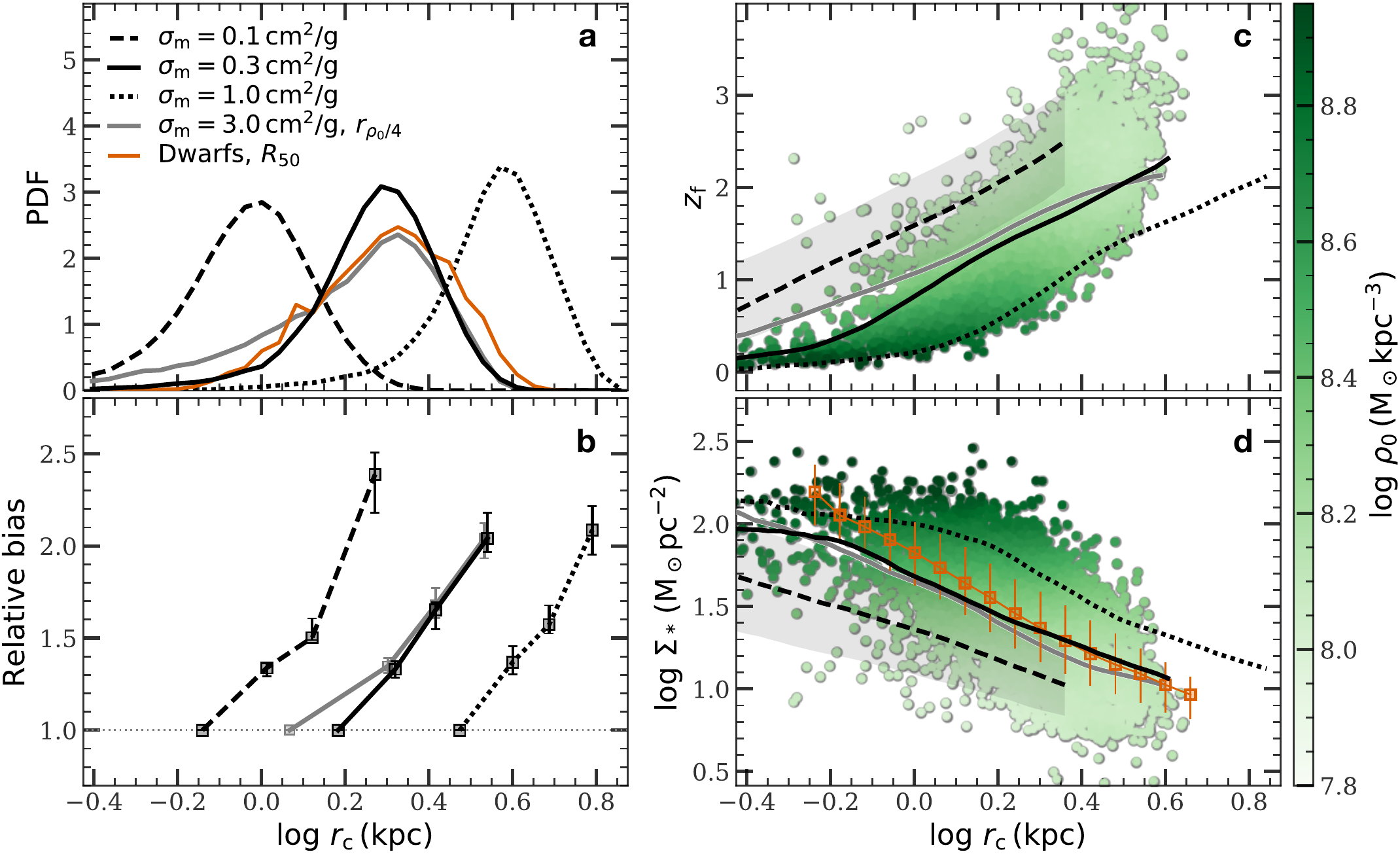}
    \caption{
    {\bf Relative bias in self-interacting dark matter (SIDM) models.} 
    Here we use an isothermal Jeans model\cite{Jiang2023} for SIDM halos, 
    adapted from the sample of CDM halos in ELUCID\cite{wangELUCIDEXPLORINGLOCAL2016} 
    with abundance-matched $\Sigma_*$ presented in 
    Fig.~\ref{fig_b_model}a and b (the $\rho=0.85$ case), to predict the core size 
    $r_{\rm c}$ and central density $\rho_0$ for each halo (see \hyperref[dis_sidm_formation]{Methods}). 
    Cases for velocity-independent cross-sections 
    $\sigma_{\rm m} = 0.1$, $0.3$ and $1.0\, {\rm cm^2\,g^{-1}}$ are shown by dashed, solid and 
    dotted black curves, respectively.
    {\bf a}, Probability density functions (PDFs) of $r_{\rm c}$.
    {\bf b}, Relative biases as functions of $r_{\rm c}$,
    binned according to the fractions of observed dwarf subsamples in
    the massive sample.
    {\bf c}, Relations between halo formation time ($z_{\rm f}$) 
    and $r_{\rm c}$.
    {\bf d}, Relations between galaxy stellar mass surface
    density ($\Sigma_*$) and $r_{\rm c}$.
    Curves with shades or error bars show medians with $16^{\rm th}$--$84^{\rm th}$
    percentiles.
    Green dots in (c) and (d) represent individual galaxies 
    for $\sigma_{\rm m}=0.3\, {\rm cm^2\,g^{-1}}$, color-coded by $\rho_0$.
    The $\sigma_{\rm m}$ in use are typical values suggested by 
    recent observational constraints\cite{Shi2021, Correa2025}.
    For comparison, results using 
    an alternative definition of core size, $r_{\rho_0/4}$, assuming 
    $\sigma_{\rm m}=3.0\, {\rm cm^2\,g^{-1}}$ are shown by grey curves.
    The distribution of $R_{50}$ and its 
    relation with $\Sigma_*$ for the observed dwarfs in the massive sample
    are shown by orange curves. Given $\sigma_{\rm m}$,
    the model predicts scaling relations 
    $\Sigma_* \propto r_{\rm c}^{-2}$ and $R_{50}\propto r_{\rm c}$ for 
    dwarfs in isolated SIDM halos.
    }
    \label{fig_sidm_bias}
\end{figure*}

Should SIDM drive the formation of diffuse dwarfs, self-interaction has to be
sufficiently strong to produce noticeable cores, thus providing testable predictions. 
We used the sample of ELUCID halos presented in Fig.~\ref{fig_b_model}
and assigned each of the halo a galaxy with $\Sigma_*$ that is obtained from its 
$z_{\rm f}$ using abundance matching. We then assumed an interacting cross-section, 
$\sigma_{\rm m}$, and adopted the isothermal Jeans model\cite{Jiang2023} 
to predict the profile (core radius, $r_{\rm c}$, and central density, $\rho_0$, defined by the expectation of ``one scattering''; see \hyperref[dis_sidm_formation]{Methods}) of SIDM. 
The result shown in Fig.~\ref{fig_sidm_bias} highlights the 
similarity between SIDM cores and dwarf galaxies, in terms of the distribution
of sizes ($r_{\rm c}$ versus $R_{\rm 50}$), and the dependencies 
of $z_{\rm f}$ and the large-scale bias on the size, indicating that the SIDM cores 
are viable proxies of structural properties of dwarf galaxies.
The predicted relation is nearly a power-law $\Sigma_* \propto r_{\rm c}^{-2}$
for a given halo mass, implying that $R_{50} \propto r_{\rm c}$ if the stellar mass 
$M_*$ in a halo depends only on the halo mass. 
Parameterizing the relation as $R_{50} = A_{\rm r} r_{\rm c}$, iterating the
Jeans model until convergence, and adjusting the normalization factor $A_{\rm r}$, 
we found that the predicted $\Sigma_*$ can reproduce the observed 
relative bias-$\Sigma_*$ relation. The model prediction and required $A_{\rm r}$ 
for given $\sigma_{\rm m}$ are shown in Fig.~\ref{fig_b_model}c and d.
For comparison, we also show in Fig.~\ref{fig_sidm_bias} the distribution of $r_{\rho_0/4}$, 
defined as the radius where the halo density drops to $\rho_0/4$\cite{Burkert95}. 
For a given cross-section $\sigma_{\rm m}$, $r_{\rho_0/4}$ is smaller than $r_{\rm c}$.
These indicates that the constraint on the cross-section depends on how $R_{50}$ is 
related to the defined core radius and can be obtained by future observations of resolved rotation curves 
for a representative population of dwarf galaxies.
Our finding clearly disfavors a large 
cross-section that leads to core collapse and inverts the trend of the bias with $\Sigma_*$.
The predicted scaling relations, $\Sigma_* \propto r_{\rm c}^{-2}$
and $R_{50}\propto r_{\rm c}$, indicated that the stellar components of diffuse 
dwarfs follow closely the dynamics driven by the dark matter. Such a condition 
may be created by a process that can effectively mix stars and star-forming gas 
with dark matter, similar to the process that produces the homology of dynamically 
hot galaxies with dark matter halos\cite{huangRelationsSizesGalaxies2017,chenTwophasePaper2-2024}.  
Clearly, these hypotheses need to be tested using hydro simulations of SIDM
that can model properly not only the dynamics of the SIDM component but 
also processes of galaxy formation. Our results provide strong motivation 
for such investigations.

\clearpage


\clearpage
\newpage

\begin{methods}\label{sec_method}

\renewcommand{\figurename}{\bf Extended Data Fig.}
\setcounter{figure}{0}
\renewcommand{\tablename}{\bf Extended Data Table}
\setcounter{table}{0}

\subsection{The sample of dwarf galaxies}\label{sec_sp}

Our galaxy sample is taken from the New York University Value Added Galaxy Catalog (NYU-VAGC)\cite{Blanton-05a} 
of the Sloan Digital Sky Survey (SDSS) DR7\cite{Abazajian-09}. We selected galaxies with the $r$-band Petrosian 
magnitudes $r\le 17.72$, the redshift completeness $\rm fgotmain\ge0.7$,  and redshift $0.01 \le z \le 0.2$. 
Isolated galaxies are defined as the central (dominating) galaxies of galaxy groups identified by the 
group-finding algorithm\cite{Yang2005, Yang2007}. The NYU-VAGC provides measurements of the size of a galaxy, 
$R_{50}$, the radius enclosing $50$ percent of the Petrosian $r$-band flux, and the $r$-band S\'ersic index, $n$. 
The $^{0.1}(g-r)$ color used here is $K$+$E$ corrected to $z=0.1$. We cross-matched the sample with the 
MPA-JHU DR7 catalog to obtain the stellar mass ($M_*$)\cite{Kauffmann2003}. As our sample of dwarf galaxies, 
we selected galaxies with $10^{7.5} \leq M_*\rm/M_{\odot} < 10^{9.0}$. 
The surface mass density of a galaxy, $\Sigma_*$, is defined as $\Sigma_* = M_*/(2\pi R_{50}^2)$. 
Extended Data Fig.~\ref{fig_para_dis}a--d shows the distributions in $^{0.1}(g-r)$, $n$, $\Sigma_*$ and $z$. 

 We excluded dwarfs with $^{0.1}(g-r)>0.6$ to reduce potential contamination  
by satellites and with $n>1.6$ to ensure a relatively pure sample of late-type galaxies.
These selections result in a sample of 6,919 galaxies (the main sample).  
As shown below, we also constructed a massive sample ($8.5<\log M_*/{\rm M_\odot}<9$ and $z\le0.04$),
which is much more complete than the main sample, and used it to compare with models.
We divide each of the main and massive samples into four subsamples according to $\Sigma_*$. 
Galaxies with $\Sigma_*<7 \, {\rm M_\odot}{\rm pc}^{-2}$ and $\Sigma_*>25 \, {\rm M_\odot}{\rm pc}^{-2}$ 
are referred to as diffuse and compact dwarfs, respectively.
Detailed information of the samples is listed in Extended Data Table~\ref{tab_gal}.
Our conclusion is robust against the details of the sample-splitting. 
The specific splitting and the $n$ cut are opted so that the diffuse dwarfs
are akin to UDGs\cite{vanDokkum2015, Koda2015, DiCintio2017, RongY2017}. 
See \hyperref[sec_sample_detail]{Supplementary Information} for details.

\subsection{The projected cross-correlation function and relative bias}\label{sec_pccf}
We first computed the two-dimensional 2PCCF using the Davis \& Peebles estimator\cite{Davis1983}. To obtain the projected 2PCCF, we integrated the 
two-dimensional one along the line of sight
within $40\, h^{-1}{\rm {Mpc}}$, sufficiently large to include almost all correlated pairs. 
The difference in redshift distribution (Extended Data Fig.~\ref{fig_para_dis}d) of 
dwarf samples necessitates a control of the redshift distributions for 
a fair comparison. 
We used two schemes, $z$-weighting and $z$-matching, to achieve this.
In the former, the diffuse sample is used as a reference, and weights are assigned to every galaxies 
in other samples to make the weighted redshift distributions the same as that for diffuse dwarfs.
In the latter, we constructed a control sample for each sample (\hyperref[sec_control_sample]{Supplementary Information}) so that all 
control samples share the same redshift distribution as indicated by the shaded region 
in Extended Data Fig.~\ref{fig_para_dis}d. Extended Data Fig.~\ref{fig_2pccf_dwarf} shows the 2PCCFs obtained using the two schemes. 

The large-scale bias is measured through the 2PCCFs. We first determined the ratio of the 2PCCF of a sample 
to that of the compact sample (Extended Data Fig.~\ref{fig_2pccf_dwarf}).
We then used a constant function $f(r_{\rm p})=b$ to model the ratio within 
$2\, h^{-1}{\rm {Mpc}}<r_{\rm p}<10\, h^{-1}{\rm {Mpc}}$ (shaded regions in Extended Data Fig.~\ref{fig_2pccf_dwarf}) and 
applied {\sc Emcee}\cite{Foreman2013} to constrain $b$. The likelihood function adopted is the same 
as equation (7) in ref.\cite{Zhang2021}, with the covariance matrix calculated as in  
ref.\cite{Trusov2023}. Extended Data Fig.~\ref{fig_2pccf_dwarf} shows results for the main sample. 
The relative bias quoted is the median of the posterior distribution, with the error 
bars indicating the $16^{\rm th}$ and $84^{\rm th}$ percentiles. 
Extended Data Table~\ref{tab_gal} shows that results from $z$-weighting and $z$-matching are similar. 
In the main text, we only show results based on the $z$-weighting scheme.

\subsection{Impacts of incompleteness, sample selection and cosmic variances}\label{sec_incomp}

The completeness of our samples can be influenced by several selection effects (SEs) 
dictated largely by the apparent magnitude and surface brightness\cite{Strauss2002}.
Given our focus on $M_*$ and $\Sigma_*$, we address the SEs in terms of $M_*$ and $\Sigma_*$. 
The apparent magnitude of a galaxy is influenced by its redshift ($z$), $M_*$, and color, 
while its surface brightness is controlled by $M_*$ and color. So the SEs are related to $z$, 
$M_*$, color, and $\Sigma_*$. The volume number densities of galaxies are directly affected by SEs 
and thus can be used to gauge their impacts. Extended Data Fig.~\ref{fig_nz} shows $n(z)$, the number density as a 
function of $z$, for different $\Sigma_*$. Since the intrinsic densities differ between different samples, 
we normalize $n(z)$ by that of the lowest-$z$ bin, $n_0$. To examine the dependence on 
$M_*$ and the color, we show results for two mass bins and two color bins. In the absence of SEs, 
$n(z)/n_0$ is expected to be roughly constant. 
A faster decline of $n(z)/n_0$ with $z$ suggests a stronger SE and thus greater incompleteness. As shown in Extended Data Fig.~\ref{fig_nz},
the SEs primarily depend on $M_*$ (or magnitude), and only weakly on $\Sigma_*$ (and size) and color for given $M_*$. 

The SEs for massive dwarfs ($8.5<\log M_*/{\rm M_\odot}<9$) at $z \le 0.04$ are much weaker than the total population.
We select these dwarfs to form a ``massive sample''. This sample is used for abundance matching which 
focuses on the $\Sigma_*$ distribution (see below). Within the massive sample, the dwarf fractions in the 
four $\Sigma_*$ bins are 2.5\%, 23.2\%, 26.0\% and 48.3\%, respectively (Extended Data Table~\ref{tab_gal}). 
These fractions at lower $z$ are similar; for example at $z\le0.03$ they are 
2.3\%, 22.2\%, 25.2\% and 50.3\%, respectively. Thus, the $\Sigma_*$ distribution 
of the massive sample is not affected significantly by the SEs.

Note that SEs should not affect the clustering strength
but only reduce the signal-to-noise ratio if they are independent of the large-scale structure (LSS). 
Thus, the SEs can affect the relative-bias measurements 
if they depend on LSS and if the dependence is different between diffuse and 
compact dwarfs. 
As shown in Strauss et al.\cite{Strauss2002}, the galaxy selection is independent of LSS, suggesting 
that the weak SEs in $\Sigma_*$ should not have significant impacts on our results. 
Since we have already controlled the redshift distribution and since the $M_*$ and color ranges 
are quite restrictive, the impact of the SEs through $M_*$ and the color is also expected to be weak.  
As a check, we made analyses in narrower ranges of redshift, mass and color (Extended Data Fig.~\ref{fig_bias_sub}a,b,c). 
If our finding was dominated by the SEs in $z$, $M_*$ or color, the trend would be weakened 
within each of the narrower bins. In contrast, the trends obtained are consistent with the results of 
the main sample. The only exception is that the relative bias of the redder sample is 
lower than the bluer one at the level of $\sim3\sigma$.
We also examined uncertainties in $M_*$, $R_{50}$ and $z$ (\hyperref[sec_uncert]{Supplementary Information}), 
and found no significant impact on our results.

Cosmic variances can have significant effects on galaxy 
statistics obtained from small samples\cite{mosterCOSMICVARIANCECOOKBOOK2011,chenELUCIDVICosmic2019}.
As shown in Extended Data Fig.~\ref{fig_bias_sub}a and d, the results obtained in distinctive volumes
defined by the two redshift intervals and the two sky areas are consistent with each other,
indicating that cosmic variances do not have big impacts on our results.

Our tests also show that 2PCCFs are highly sensitive to contamination by satellite galaxies only 
on small scales (Extended Data Fig.~\ref{fig_2pf}), indicating that the contamination  cannot explain our 
finding that is based on large-scale clustering.

To test the impact of the cut in S\'ersic index used in our sample selection, 
we conducted test by removing the cut (Extended Data Fig.~\ref{fig_bias_sub}e).
Without restricting $n$, diffuse dwarfs are still more strongly clustered than compact dwarfs. 
However, since the dependence on $\Sigma_*$ is weak for dwarfs with $n>1.6$ 
(Extended Data Fig.~\ref{fig_bias_sub}e), including large-$n$ dwarfs weakens the $\Sigma_*$ dependence. 
Extended Data Fig.~\ref{fig_bias_sub}f shows that the relative bias is quite independent of $n$, 
indicating that the assembly bias is not well reflected by $n$. 
Apparently, the $\Sigma_*$ of large-$n$ dwarfs are not determined by halo assembly 
history, in contrast to that of small-$n$ dwarfs, but the physics behind it is not yet understood.  
Including large-$n$ dwarfs thus dilutes the signal of assembly bias and 
complicates the interpretation of results. Because of this, we excluded dwarfs with $n>1.6$ 
(about $28\%$ of the total) from the main sample.  

\subsection{Halo mass estimates}\label{sec_shmr}
 The halo mass is defined as the mass enclosed by 
the radius within which the mean density is 200 times the mean matter density of the Universe at the epoch in question.
We first adopted the SHMR\cite{Kravtsov2018} to estimate the halo mass. 
Abundance matching found that scatter in $M_*$, $\sigma_{\log M_*}$,
is about $0.2 \, {\rm dex}$ at given $M_{\rm h}$\cite{Wechsler2018}. Thus, the scatter in $M_{\rm h}$ is $\sigma_{\log M_{\rm h}}=\sigma_{\log M_*}\frac{{\rm d}(\log M_{\rm h})}{{\rm d}(\log M_{*})}\sim0.1$ at $\log M_*/{\rm M_\odot}\sim9$.
We estimated the median $M_{\rm h}$ and its uncertainty for a sample as follows. 
For a given galaxy, the uncertainty in $M_*$ is considered to be Gaussian, 
with a spread set by the measurement error of $M_*$. A random stellar mass, $M_{\rm *,r}$, 
is assigned to the galaxy. 
Halo mass at given stellar mass is also assumed to follow a Gaussian distribution with a dispersion of $0.1\, {\rm dex}$.
We then generated a random halo mass from $M_{\rm *,r}$ and used the halo bias model\cite{Tinker2010} to predict a halo bias. Finally, we obtained one measurement of the median 
$M_{\rm h}$ and the mean bias of the sample. This process was repeated $100$ times, 
yielding $100$ measurements of the median $M_{\rm h}$ and the mean bias. 
The $50^{\rm th}$ percentiles of these measurements represent the median halo mass
and halo bias for the sample, while the $16^{\rm th}$ and $84^{\rm th}$ percentiles represent 
their uncertainties (as listed in Extended Data Table~\ref{tab_gal}). 
The predicted bias ratio between diffuse and compact dwarfs is 
$0.99$ with an uncertainty less than $0.01$.

We then used HI kinematics to measure the halo mass. We cross-matched our dwarf sample with the complete 
Arecibo Legacy Fast Arecibo L-band Feed Array (ALFALFA $\alpha.100$) HI survey\cite{Giovanelli2005,Haynes2018}. To estimate the rotation 
velocities and halo masses, we excluded galaxies with dubious HI spectra, low HI spectra signal-to-noise ratios 
(SNR$<8$) and large axis ratios ($b/a>0.7$). We used the same method outlined in ref.\cite{Guo2020-UDG} to obtain 
the rotation velocity from the line width ($W_{20}$) and the halo mass by assuming the Burkert profile\cite{Burkert95} 
with a central core\cite{Marchesini02,Rong24a} (see also \hyperref[sec_detail_of_HI_mh]{Supplementary Information}). The halo mass uncertainty is 
determined by taking into account the uncertainties in stellar mass, HI mass, HI line width, inclination and the 
assumed profile ($\sim 0.15\, {\rm dex}$\cite{Wangj20}). Since resolved HI maps 
are unavailable, we used the inclination of the stellar disk\cite{Guo2020-UDG}
to estimate the HI inclination, assuming a misalignment given by a Gaussian distribution with 
dispersion ${\rm \delta}\phi \simeq 20^\circ$\cite{Starkenburg-counterrotation, Guo2020-UDG, Gault2021-VLA_UDG}. 

Extended Data Fig.~\ref{fig_MhHI} shows the halo mass obtained from HI kinematics,
$M_{\rm h, HI}$, versus $M_*$. The overall trends in the $M_{\rm h, HI}$-$M_*$ relations 
resemble the SHMR\cite{Kravtsov2018}, but with much larger dispersion due to the uncertainties 
in $M_{\rm h, HI}$. Our estimates for diffuse dwarfs are consistent with those 
of UDGs obtained by ref.\cite{Kong2022ApJ} from HI rotation curves.
The uncertainty of individual galaxies surpasses the $M_{\rm h,HI}$ measurement dispersion, and is thus overestimated, primarily due to the inclination errors.
Assuming the uncertainty of $M_{\rm h, HI}$ to follow a Gaussian with dispersion
equal to its error, we generated a new $M_{\rm h, HI}$ and predicted a bias 
$b(M_{\rm h, HI})$\cite{Tinker2010} for each galaxy. We then adopted the same method as for the SHMR mass 
to obtain the median halo mass/halo bias and their uncertainties for individual samples(Extended Data Table~\ref{tab_gal}).
The predicted bias ratio between the diffuse and compact samples is $0.66/0.7=0.94$, with
an uncertainty $\sim0.02$.

\subsection{HI mass of dwarf galaxies}\label{sec_HImass}
We cross-matched the optical 
counterparts of the ALFALFA sample\cite{Giovanelli2005,Haynes2018} with our dwarf galaxies.
The HI detection rates for the four samples in the ascending order of $\Sigma_*$ 
are $84.0\%$, $68.1\%$, $49.6\%$ and $35.6\%$, respectively.
Extended Data Fig.~\ref{fig_HIm} shows the HI mass for galaxies with HI detections. Clearly, diffuse dwarfs are gas richer than compact ones, 
suggesting that they cannot be produced by environmental processes capable of stripping their extended HI disks.

\subsection{The distribution of dwarf galaxies in the cosmic web}\label{sec_cosmic_web}
To investigate the connection between dwarf galaxies and the cosmic web, 
we used the reconstructed mass density field of the local Universe provided by
the ELUCID project\cite{wangELUCIDEXPLORINGLOCAL2016}.
The cosmic web was classified using the ``T-Web'' method\cite{Hahn2007}, which utilizes eigenvalues of 
the local tidal tensor to define the morphology of the local structure as knot, filament, sheet and void. 
The grey shades in Fig.~\ref{fig_cosmicweb}c and d show the fraction of filament grids along each line-of-sight. Since the redshift-space distortion (RSD) is corrected in the reconstruction, we assigned a corrected redshift\cite{wangELUCIDEXPLORINGLOCAL2016}, $z_{\rm cor}$, to each of the galaxies and groups shown in Fig.~\ref{fig_cosmicweb}a--d.
To quantify the spatial correlation between dwarfs and the cosmic web, we computed
the 2PCCF in real space between dwarf galaxies in our main sample
and different grid points. 
Galaxies are $z$-weighted to match the 
redshift distribution of diffuse dwarfs, and grid points 
are weighted by their matter density. 
Fig.~\ref{fig_cosmicweb}e and f show the eight 2PCCFs, highlighting the difference in
large-scale environment between diffuse and compact dwarfs. We calculated the projected distance from a 
diffuse dwarf to the nearest group (see \hyperref[sec_distance_to_group]{Supplementary Information}) and found that the median distance is 
significantly higher than that for backsplash halos\cite{Wang2009}, indicating that diffuse dwarfs are not 
backsplashs. This also aligns with the observation that diffuse dwarfs contain more HI-gas than compact dwarfs.

\subsection{Halo assembly bias in cosmological simulations}\label{sec_bias_in_sim}

We analyzed the dark-matter-only (DMO) simulation, TNG300-1-Dark\cite{nelsonIllustrisTNGSimulationsPublic2019},
to explore whether halo assembly bias\cite{Gao2005} can explain the observed bias-$\Sigma_*$ relation. 
The resolution of this simulation allows us to compute halo spin accurately. 
We excluded backsplash halos, as they are unlikely to be relevant to diffuse dwarfs.  

Halos with $10^{10.5} \leq M_{\rm h}/{\rm M_\odot} < 10^{11}$ were divided into 
subsamples by half-mass formation time\cite{liHaloFormationTimes2008}  ($z_{\rm f}$)
or spin\cite{bullockUniversalAngularMomentum2001} ($\lambda$). 
The reference sample to estimate the 2PCCF included all centrals and satellites 
with $M_{\rm h,peak} \geq 10^{10.5}\, {\rm M_\odot}$, where $M_{\rm h,peak}$ denotes peak 
main-branch halo mass. We incorporated redshift-space distortions (RSD) along one simulation axis\cite{chenELUCIDVICosmic2019}. 
Extended Data Fig.~\ref{fig_2pccf_model}a and b show that dwarf-host halos with the 
highest $z_{\rm f}$ have clustering comparable to halos with $M_{\rm h} \gtrsim 10^{13}\, {\rm M_\odot}$.  

Our findings imply that the bias-$z_{\rm f}$ relation can explain the observed bias-$\Sigma_*$ relation, 
provided that $z_{\rm f}$ governs $\Sigma_*$. 
To see this, we applied an abundance matching\cite{Hearin2013agematching} 
between $\Sigma_*$ in the massive sample and $z_{\rm f}$ of dwarf-host halos 
in the same volume simulated by ELUCID\cite{wangELUCIDEXPLORINGLOCAL2016}, 
assuming some scatter in the matching. ELUCID is an N-body simulation constrained to reproduce the 
density field underlying SDSS galaxies, thus ensuring the same large-scale environments 
for the simulated halos and observed dwarfs. The $\Sigma_*$–$z_{\rm f}$ mapping follows\cite{Behroozi2019} 
\begin{equation}
    \Sigma_* = 
    \mathcal{P}_{\rm \Sigma_*}^{-1} \circ \mathcal{N} \left[
        - \rho \mathcal{N}^{-1} \circ \mathcal{P}_{z_{\rm f}}(z_{\rm f})
        + \sqrt{1 - \rho^2} \epsilon
    \right]\,,
\end{equation}  
where $\mathcal{N}$ is the cumulative distribution function (CDF) of 
a Gaussian variable;
$\mathcal{P}_{z_{\rm f}}$ and $\mathcal{P}_{\Sigma_*}$ are CDFs of
$z_{\rm f}$ and $\Sigma_*$, respectively, obtained numerically 
from the samples in question;
``$\circ$'' denotes function composition and 
``$^{-1}$'' denotes functional inversion;
$\rho$ quantifies the $z_{\rm f}$–$\Sigma_*$ correlation 
and $\epsilon$ is a unit Gaussian random noise.
This matching assigns a $\Sigma_*$ to each halo, preserving the 
observed $\Sigma_*$ distribution. The relative bias of halos
is shown in Fig.~\ref{fig_b_model}a as a function of the assigned $\Sigma_*$,
assuming different $\rho$. See \hyperref[sec_AM]{Supplementary Information} for more
details of abundance matching.

\subsection{The formation of diffuse dwarfs in the cold dark matter scenario}\label{sec_dis_dd_formation}

Current models for the formation of (ultra-)diffuse dwarfs in CDM halos  
fail to reproduce the observed bias-$\Sigma_*$ relation. 
Tidal heating\cite{Safarzadeh2017,Jiang2019}, galaxy interactions\cite{Silk2019}, 
and ram pressure stripping\cite{Benitez-Llambay2013} require dense environments 
(groups/filaments) which remove gas or quench star formation, incompatible 
with the blue, HI-rich nature of diffuse dwarfs(Extended Data Fig.~\ref{fig_HIm}). 
Models attributing diffuse dwarfs to suppressed star formation in massive halos interacting with 
the large-scale structure\cite{Yozin2015,vanDokkum2016} conflict with the halo-mass estimates and 
the small-scale clustering (Fig.~\ref{fig_bias}a). 
Models relying on exceedingly high-spin halos to host diffuse dwarfs
\cite{Amorisco2016,RongY2017,Benavides2023} 
predict a bias-$\Sigma_*$ relation that is inconsistent with observations (see 
Fig.~\ref{fig_b_model} for L-Galaxies), as the assembly bias in halo spin is too weak 
(Extended Data Fig.~\ref{fig_2pccf_model}).  
Episodic stellar/supernova feedback-driven outflows and associated
variations of gravitational potential, seen in simulations like NIHAO\cite{DiCintio2017} 
and FIRE\cite{Chan2018}, could cause galaxies and halos to expand. 
However, these models disfavor UDGs in halos of high concentration (thus high-$z_{\rm f}$\cite{chenRelatingStructureDark2020}, high-bias; see Extended Data Fig.~\ref{fig_2pccf_model}c), and cause deficits of compact dwarfs\cite{Jiang2019} and 
steep dark-matter profiles\cite{Relatores2019}. 

To demonstrate the discrepancy between the models and our observation,
we directly compared our results with two models, the TNG100-1 hydro simulation\cite{pillepichFirstResultsIllustrisTNG2018,
Springel2018,nelsonFirstResultsIllustrisTNG2018,naimanFirstResultsIllustrisTNG2018,
marinacciFirstResultsIllustrisTNG2018,
nelsonIllustrisTNGSimulationsPublic2019,weinbergerSimulatingGalaxyFormation2017,
pillepichSimulatingGalaxyFormation2018}
and the L-Galaxies semi-analytic model\cite{henriquesGalaxyFormationPlanck2015,ayromlouComparingGalaxyFormation2021}. 
Central star-forming dwarfs in both models show weak $z_{\rm f}$-$\Sigma_*$ relations (Fig.~\ref{fig_b_model}b)
and have 2PCCFs (Extended Data Fig.~\ref{fig_2pccf_model}d and e) that are inconsistent with the observed $\Sigma_*$ dependence.
Tests using TNG with higher resolutions\cite{pillepichFirstResultsTNG502019,nelsonFirstResultsTNG502019,nelsonIllustrisTNGSimulationsPublic2019}
(Extended Data Fig.~\ref{fig_2pccf_model}f) and L-Galaxies in a larger volume proved that our conclusions are 
robust. We also found that backsplash halos have negligible effects on the large-scale bias.  
We suspect that the discrepancy arises from model assumptions: L-Galaxies ties 
cold-gas sizes to halo spins\cite{guoDwarfSpheroidalsCD2011} (anti-correlated with 
$z_{\rm f}$\cite{chenRelatingStructureDark2020}), while in TNG the sizes are regulated by 
stellar winds\cite{pillepichSimulatingGalaxyFormation2018} that may erase halo assembly effects.

\subsection{Assembly bias in self-interacting dark matter models}\label{dis_sidm_formation}

Dark matter self-interaction flattens halo central profiles while preserving 
the outer shape and large-scale clustering. 
The thermalized SIDM core can be described by its central density ($\rho_0$) and core
radius ($r_{\rm c}$) at which each particle is expected to experience one scattering over the halo 
lifetime\cite{Kaplinghat2016}, both governed by the cross-section per particle mass, 
$\sigma_{\rm m}$. Alternative definitions of the core size also exist, 
e.g., $r_{\rho_0/4}$, defined as the radius at which the density drops to 
$\rho_0/4$\cite{Burkert95}. 

Large SIDM simulations capable of resolving halos of dwarfs remain impractical. 
We instead applied a semi-analytical 
method\cite{Kaplinghat2016,Jiang2023} to CDM halos used in Fig.~\ref{fig_b_model}a 
and b to predict SIDM cores via the isothermal Jeans modeling. 
Concentrations of ELUCID halos were assigned using the conditional distribution
$p(c | z_{\rm f})$ calibrated from a simulation with higher resolution. 
Each halo is populated with a galaxy of $M_*=10^{8.8}\, {\rm M_\odot}$ (Extended Data Table~\ref{tab_gal}) 
and an exponential profile according to its $\Sigma_*$ assigned by the abundance matching. 
Adiabatic contraction due to baryons and Jeans modeling\cite{Jiang2023} were then applied
to predict $r_{\rm c}$, $r_{\rho_0/4}$ and $\rho_0$.  

Current constraints on $\sigma_{\rm m}$ for low-mass halos range 
from $\leq 1.63\, {\rm cm^2\,g^{-1}}$ (based on inner halo profiles)\cite{Shi2021} 
to $\leq10\, {\rm cm^2\,g^{-1}}$ (based on the Tully-Fisher relation)\cite{MoMao2000,Correa2025}.
See refs.\cite{Correa2025,Fischer2024} for a summary. For demonstration 
we adopted velocity-independent $\sigma_{\rm m}=0.1$–$1.0\, {\rm cm^2\,g^{-1}}$ for $r_{\rm c}$ and $\rho_0$ and 
$3.0\, {\rm cm^2\,g^{-1}}$ for $r_{\rho_0/4}$. Fig.~\ref{fig_sidm_bias} shows that (i) 
the $r_{\rm c}$ distribution assuming $\sigma_{\rm m}=0.3\, {\rm cm^2\,g^{-1}}$ 
aligns with the observed $R_{50}$ distribution, with a higher $\sigma_{\rm m}$ 
predicting a proportionally shifted distribution to the right (panel a);
(ii) the tight monotonic bias-$r_{\rm c}$ (panel b) and $z_{\rm f}$-$r_{\rm c}$ (panel c) 
relations mirror the observed bias-$\Sigma_*$ ($R_{50}$) relation (Fig.~\ref{fig_b_model});
(iii) the $\Sigma_*$-$r_{\rm c}$ and $\Sigma_*$-$R_{50}$ relations match each other 
closely (panel d); (iv) matching $R_{50}$ with $r_{\rho_0/4}$ requires larger $\sigma_{\rm m}$. 
We also found that the inclusion of baryons in the adiabatic contraction and in the Jeans-Poisson 
equation makes the core size larger, more so for halos with lower $z_{\rm f}$, 
but does not disorder the $r_{\rm c}$ (or $r_{\rho_0/4}$) - $R_{50}$
relation, provided that $\sigma_{\rm m}$ is not so large that core collapse inverts the 
bias-$\Sigma_*$ relation required by the observation. Note that our predictions 
for SIDM cores rely on the sequence of assumptions that were incrementally 
incorporated. See \hyperref[sec_model_assumptions]{Supplementary Information} for more details.

\end{methods}

\begin{addendum}
\item[Data availability]
The stellar mass and star formation rate for SDSS galaxies used in this paper are publicly available at \url{https://wwwmpa.mpa-garching.mpg.de/SDSS/DR7/}. 
The galaxy size and S\'ersic index data can be downloaded at \url{http://sdss.physics.nyu.edu/vagc/}.
The galaxy group catalog is publicly available at \url{https://gax.sjtu.edu.cn/data/Group.html}. 
The ALFALFA HI sample can be downloaded at \url{https://egg.astro.cornell.edu/alfalfa/data/}.
The simulation data are available through the IllustrisTNG public data release\cite{nelsonIllustrisTNGSimulationsPublic2019} at \url{https://www.tng-project.org/}
for the runs used in this paper, and for L-Galaxies implemented on the runs.
The ELUCID simulation data are available upon request.

\item[Code availability]
The code used in this paper is available at \url{https://github.com/ChenYangyao/dwarf_assembly_bias}.
The code for the semi-analytic method based on the isothermal Jeans model is publicly available at \url{https://github.com/JiangFangzhou/SIDM}.
\end{addendum}

%% Here is the endmatter stuff: Supplementary Info, etc.
%% Use \item's to separate, default label is "Acknowledgements"

\begin{addendum}
 \item[Acknowledgements] 
We thank Prof. Ethan O. Nadler and the two anonymous referees for the useful reports that significantly improve this paper. 
We thank Fangzhou Jiang, Liang Gao, Jie Wang, Qi Guo, Xiaohu Yang, Ying Zu, Ran Li, Daneng Yang, Ce Gao and Kai Wang for comments.
This work is supported by the National Natural Science Foundation of China (NSFC, Nos. 12192224, 12273037 and 11890693). HYW acknowledgements supports from CAS Project for Young Scientists in Basic Research, Grant No. YSBR-062. 
 Y.R. acknowledgements supports from the CAS Pioneer Hundred Talents Program (Category B), as well as the USTC Research Funds of the Double First-Class Initiative.
Y.C. acknowledgements supports from China Postdoctoral Science Foundation (Grant No. 2022TQ0329).
We acknowledge the science research grants from the China Manned Space Project with CMS-CSST-2021-A03 and Cyrus Chun Ying Tang Foundations.  The work is also supported by the Supercomputer Center of University of Science and Technology of China, and the Tsinghua Astrophysics High-Performance Computing platform of Tsinghua University. HYW acknowledges the hospitality of the International Centre of Supernovae (ICESUN), Yunnan Key Laboratory at Yunnan Observatories Chinese Academy of Sciences.
\item[Author Contributions] The listed authors made substantial contributions to this manuscript; all co-authors read
and commented on the document. ZWZ, YYC and YR contributed equally to this work, and YYC and YR are co-first 
authors of this paper. HYW conceived the original idea, initiated the project and led the analysis.
HJM contributed to the writing and interpretation of results. XL contributed to the analysis of observational data. HL contributed to the analysis of simulation data.
 \item[Competing Interests] The authors declare that they have no competing financial interests.
 \item[Supplementary Information] is available for this paper.
 \item[Correspondence] Correspondence and requests for materials should be addressed to HYW~(email: whywang@ustc.edu.cn). 
 \item[Reprints and permissions information] is available at www.nature.com/reprints.
\end{addendum}

\clearpage

\renewcommand{\figurename}{\bf Extended Data Fig.}
\setcounter{figure}{0}
\renewcommand{\tablename}{\bf Extended Data Table}
\setcounter{table}{0}

\begin{figure}
    \centering
    \includegraphics[scale=0.3]{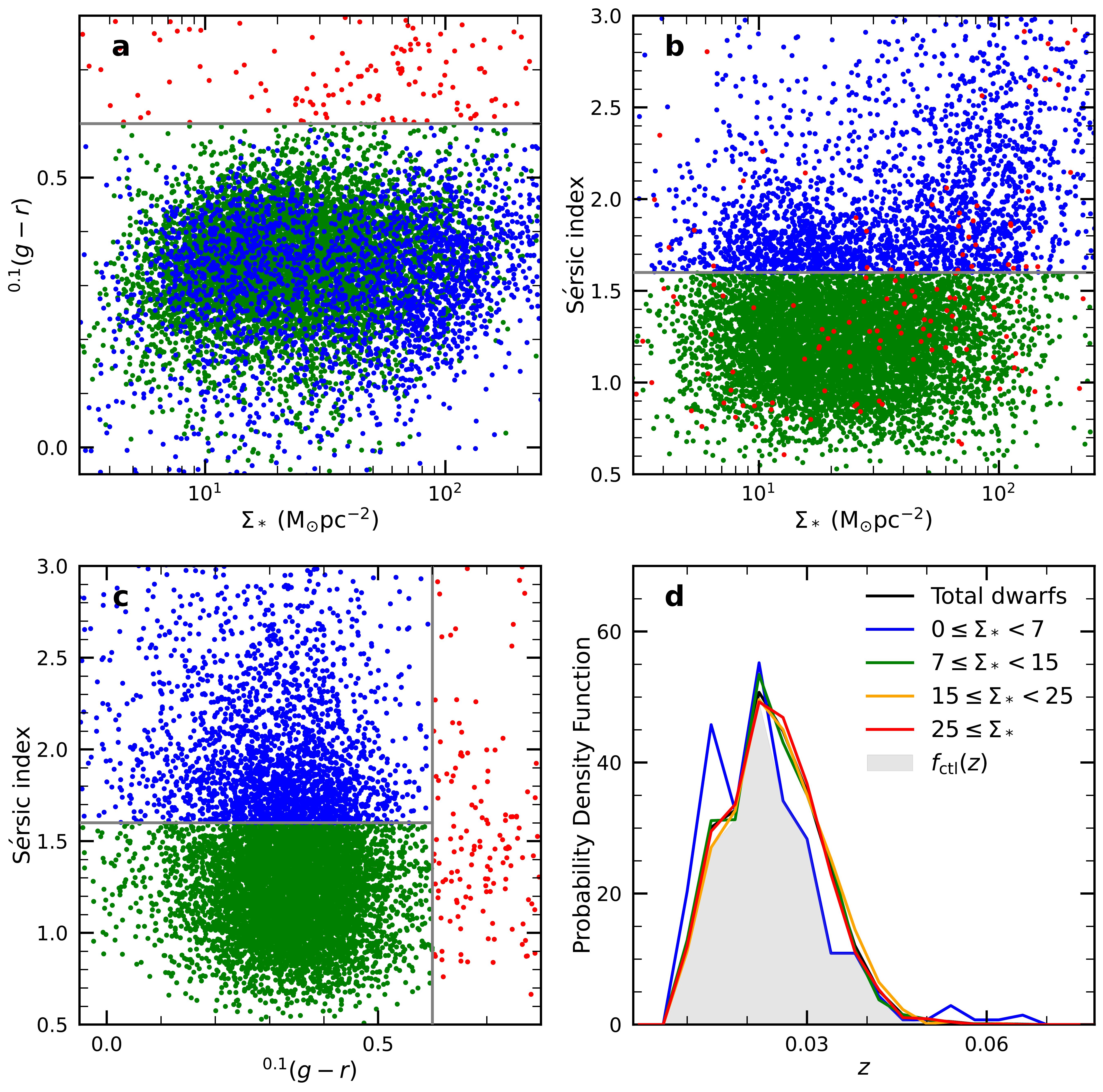}
    \caption{{\bf Dwarf galaxy sample selection}. {\bf a, b, c}, Distributions
    of dwarf properties. Green dots represent the finally selected dwarfs. 
    Red dots show the dwarfs with $^{0.1}(g-r)>0.6$ and 
    blue dots show the dwarfs with $^{0.1}(g-r)<0.6$ and $n>1.6$. 
    {\bf d}, Redshift distributions of dwarf galaxy samples with 
    different $\Sigma_*$. 
    Shaded region shows the redshift distribution for control samples, 
    $f_{\rm ctl}(z)$, used in the $z$-matching method.}
    \label{fig_para_dis}
\end{figure}

\begin{figure}
    \centering
    \includegraphics[scale=0.25]{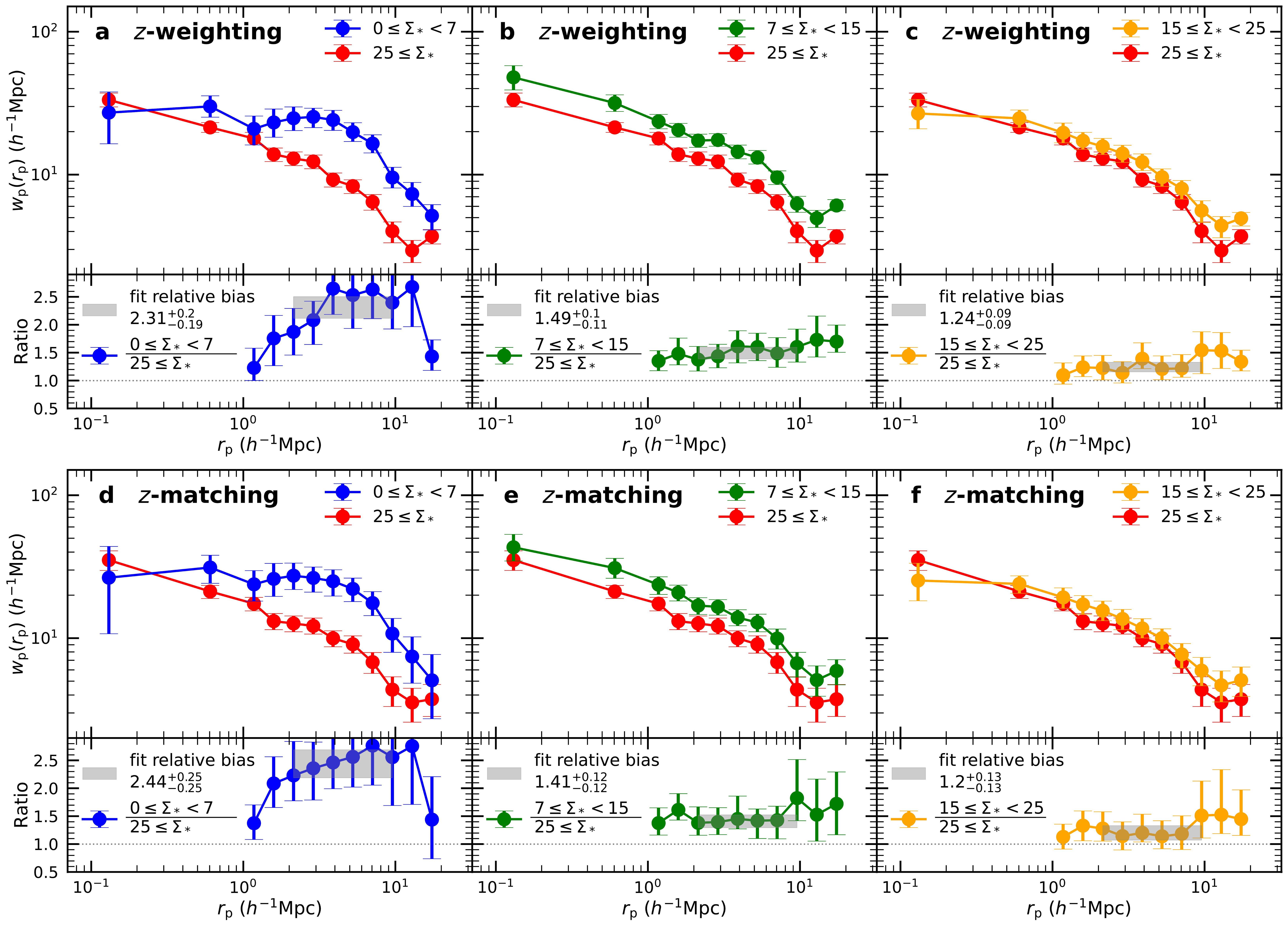}
    \caption{{\bf 2PCCFs for the main samples.} The first and third rows show the 2PCCFs for dwarfs with different surface density. And the second and fourth rows show the 2PCCF ratios relative to compact dwarfs. The top two rows show the results using the $z$-weighting method, while the bottom two rows present those for the $z$-matching method. The error bars for both the 2PCCFs and the 2PCCF ratios represent the $16^{\rm th}$ and $84^{\rm th}$ percentiles of 100 bootstrap samples. 
    The shaded region indicates the radius interval used for fitting and best-fit relative bias.
    The error bars for relative bias represent the $16^{\rm th}$ and $84^{\rm th}$ percentiles of the posterior distribution.}
    \label{fig_2pccf_dwarf}
\end{figure}

\begin{figure}
    \centering
    \includegraphics[scale=0.3]{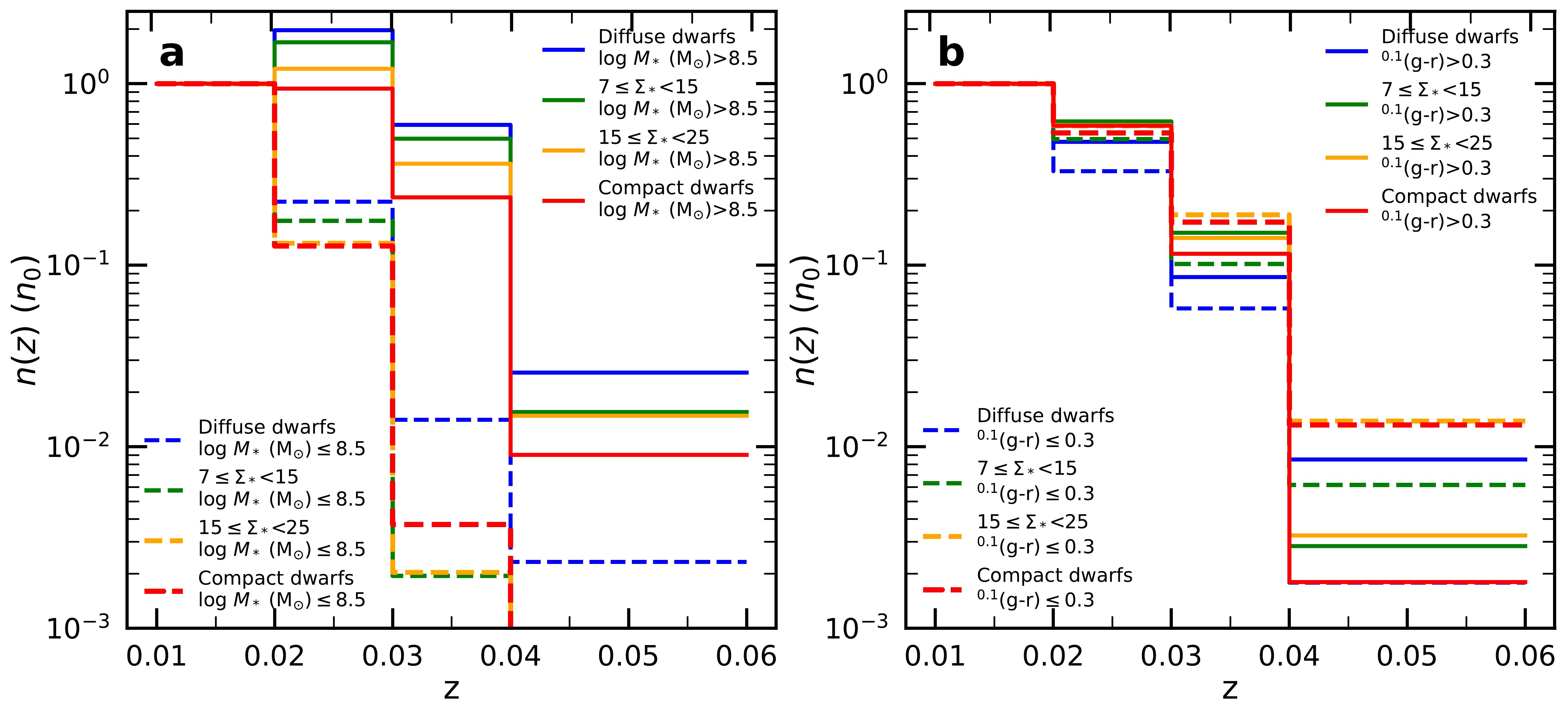}
    \caption{{\bf Number density $n(z)$ as a function of redshift for different $\Sigma_*$.} 
    $n(z)$ is normalized by that of the lowest-$z$ bin ($n_0$). 
    {\bf a}, $n(z)$ for low-mass ($7.5<\log M_*/{\rm M_\odot}\leq8.5$) and 
    massive ($8.5<\log M_*/{\rm M_\odot}\leq9$) dwarfs separately. 
    For massive dwarfs, the SEs become large only when $z>0.04$. 
    For less-massive dwarfs, the SEs are significant even at $z\sim0.02$. 
    For given $M_*$, the impact of the SEs depends only weakly on $\Sigma_*$, 
    as is expected from the small redshift concerned here. 
    At $z>0.04$, there is no low-mass dwarf with $\Sigma_*>7\,{\rm M_\odot}{\rm pc}^{-2}$. 
    {\bf b}, $n(z)$ for red ($0.3<^{0.1}(g-r)<0.6$) and blue ($^{0.1}(g-r)<0.3$) 
    dwarfs separately. Dwarfs with different colors exhibit similar behavior, 
    indicating that the SEs are insensitive to galaxy color. 
    This is because our galaxies have already been restricted to a relatively 
    narrow color range.  }
    \label{fig_nz}
\end{figure}

\begin{figure}
    \centering
    \includegraphics[scale=0.2]{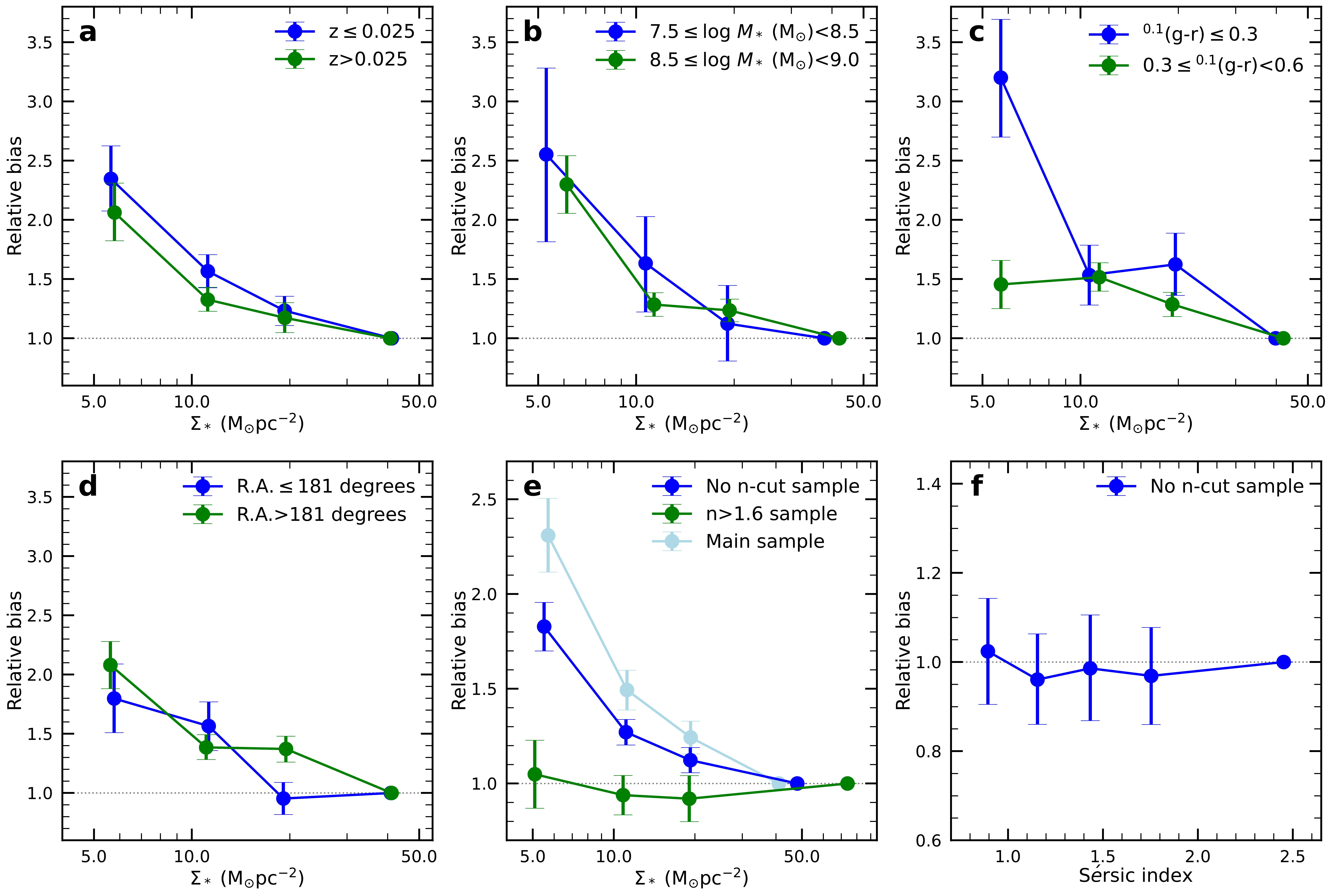}
    \caption{{\bf Relative biases obtained based on different dwarf subsamples.} 
    Here the samples are divided by $z$ ({\bf a}), 
    $M_*$ ({\bf b}), color ({\bf c}), Right Ascension (R.A., {\bf d}),
    and S\'ersic $n$ ({\bf e}), respectively, and the relative biases 
    versus $\Sigma_*$ are shown for subsamples.
    In {\bf e}, the main sample is exactly the sample used in the main text. 
    The $n>1.6$ sample consists of isolated dwarf galaxies with $n>1.6$ and 
    $^{0.1}(g-r)<0.6$. 
    The no $n$-cut sample includes the main sample and $n>1.6$ sample. 
    Note that the three curves are normalized to different compact samples that 
    may have different clustering strength.
    {\bf f}, Relative bias as a function of $n$ for no $n$-cut dwarf sample. 
    The relative bias is normalized to the subsample with the largest $n$. 
    Only results using the $z$-weighting method are shown here. 
    The results from the $z$-matching method are very similar and thus not presented. 
    Markers with error bars are median values with $16^{\rm th}$--$84^{\rm th}$ 
    percentiles of relative biases obtained from the posterior distribution 
    of MCMC fitting.}
    \label{fig_bias_sub}
\end{figure}

\begin{figure}
\includegraphics[scale=0.4]{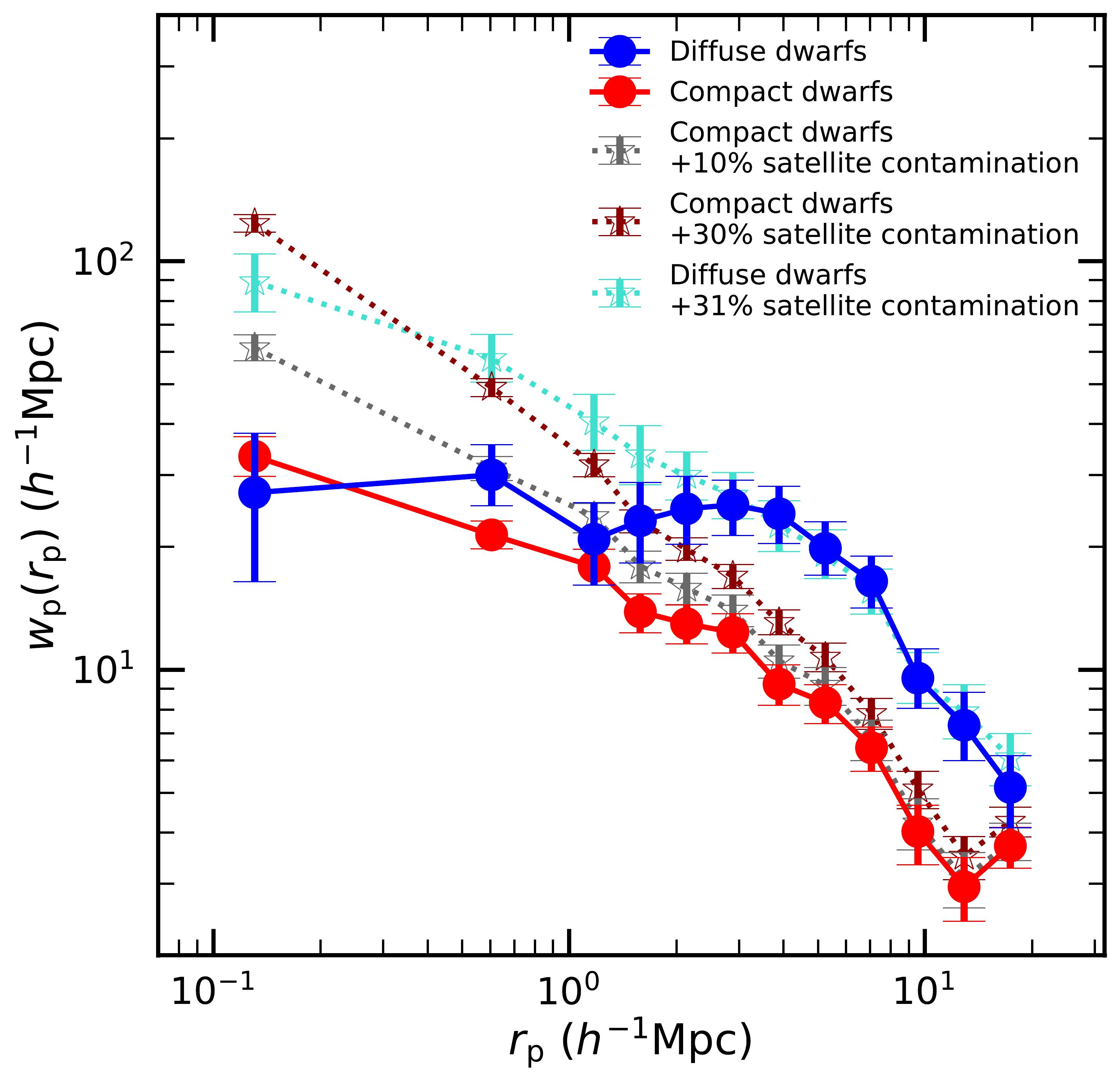}
\centering
\caption{{\bf 2PCCFs with satellite contamination.}
    Blue and red solid curves represent the 2PCCFs for diffuse and compact dwarf galaxies, 
    respectively, while dotted curves show the impact of different levels of satellite contamination on these dwarfs. Satellite contamination can notably amplify small-scale clustering, while it moderately enhances large-scale clustering for compact dwarfs and leaves the large-scale clustering unchanged for diffuse dwarfs. 
    Note that the wine and cyan dotted lines show the results including all 
    compact and diffuse satellite dwarfs, respectively. Thus, satellite contamination 
    cannot explain the strong large-scale clustering observed in isolated diffuse dwarfs. 
    Error bars represent $16^{\rm th}$--$84^{\rm th}$ percentiles of bootstrap 
    samples.}
    \label{fig_2pf}
\end{figure}

\begin{figure}
    \centering
    \includegraphics[scale=0.3]{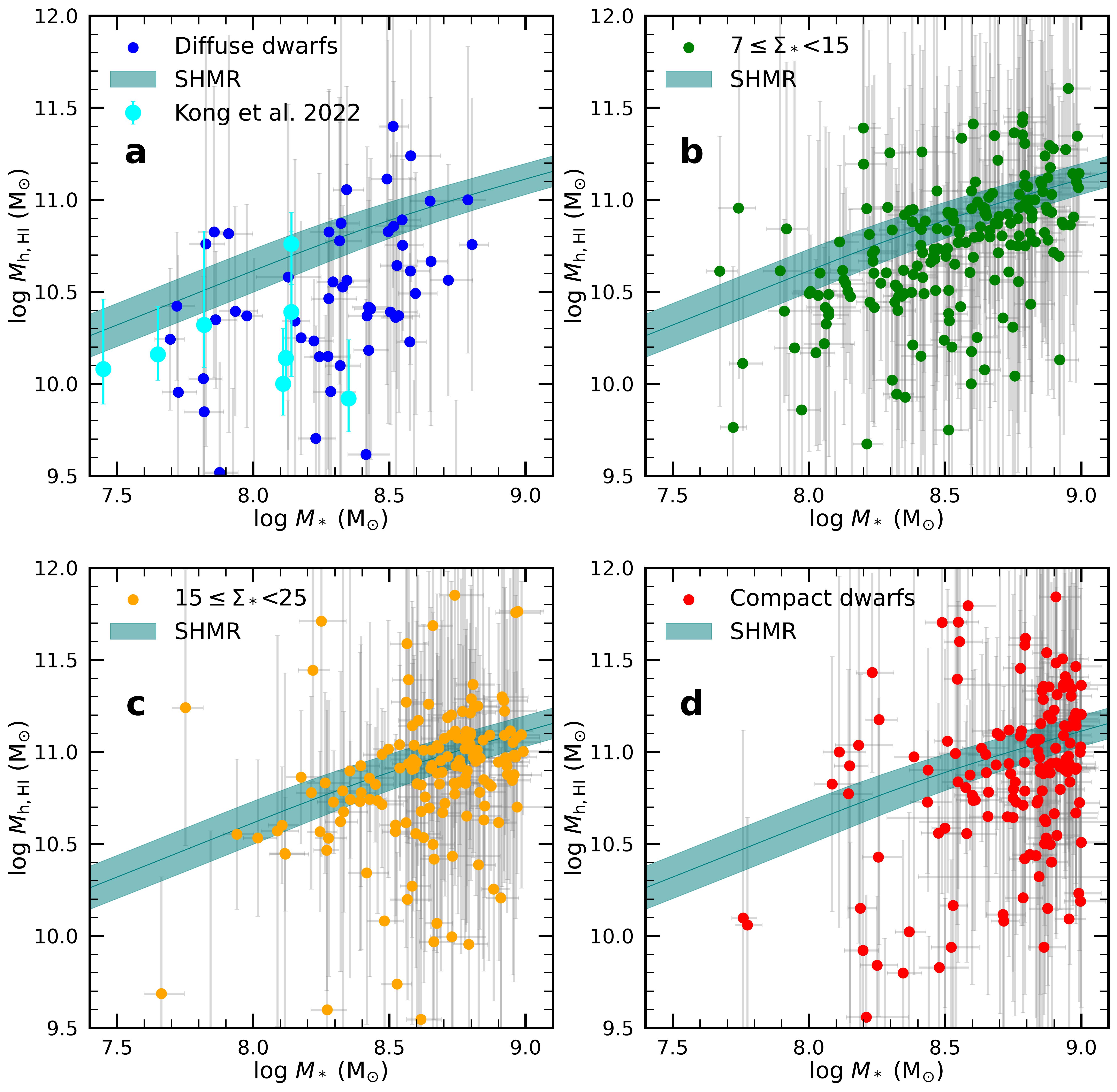}
    \caption{{\bf Comparison of halo masses of dwarf galaxies derived from different methods.} {\bf a--d}, halo mass versus $M_*$ for dwarf samples
    with different $\Sigma_*$.
    Symbols with error bars show the halo mass obtained by using the HI 
    kinematics versus $M_*$ and their uncertainties.
    Teal shadow region shows the SHMR\cite{Kravtsov2018} and its 1$\sigma$ 
    uncertainty. Cyan symbols show the results for UDGs taken from 
    ref.\cite{Kong2022ApJ}. These UDGs have spatially-resolved HI kinematics maps, therefore their halo mass measurements are more reliable than ours. As can be seen, these UDGs follow the same trend as our diffuse dwarfs.}
    \label{fig_MhHI}
\end{figure}

\begin{figure*} \centering
    \includegraphics[width=\textwidth]{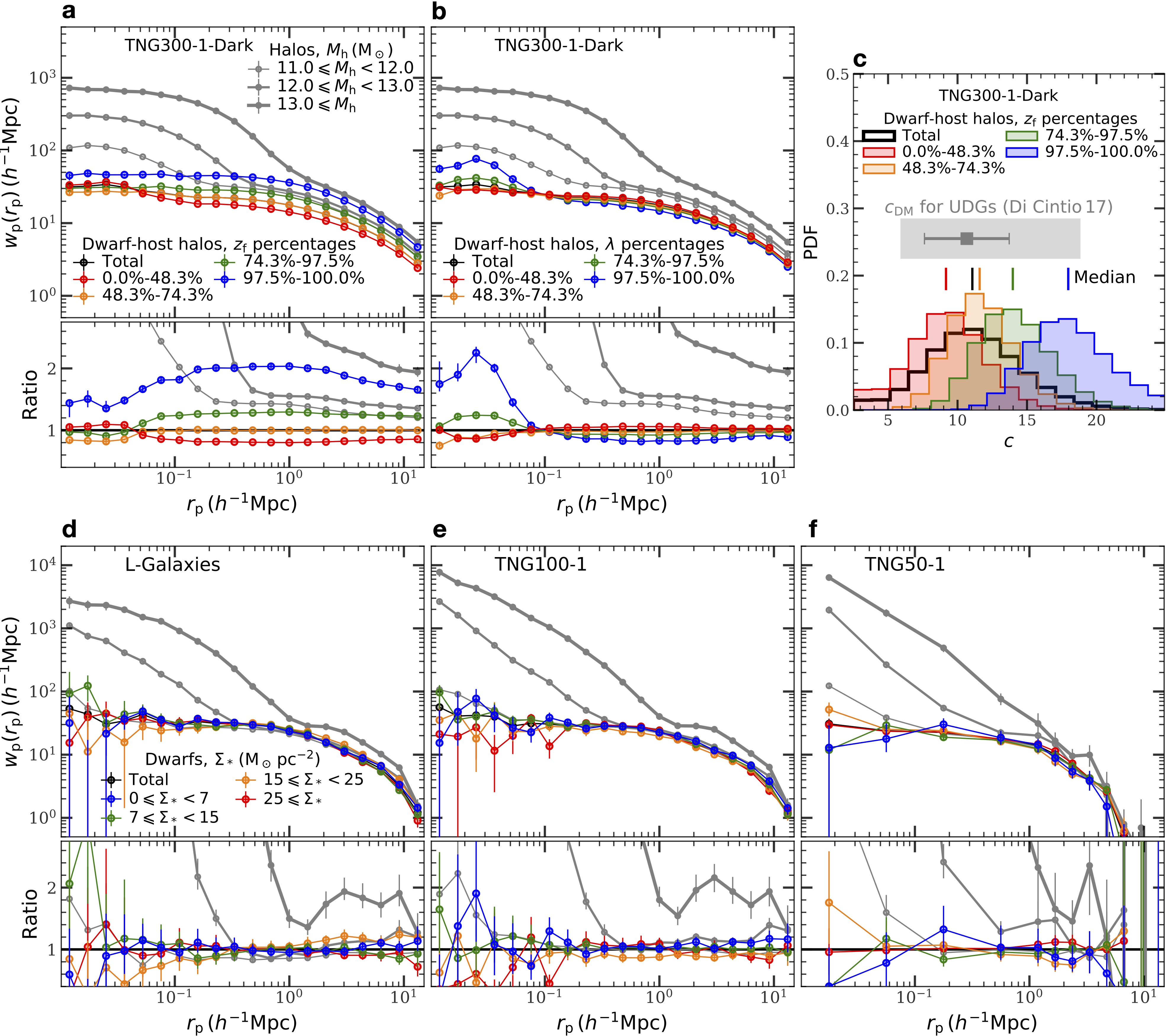} 
    \caption{
    {\bf Numerical simulations for dwarf galaxies and dwarf-host halos at $z=0$.}
    {\bf a, b}, 2PCCF of dwarf-host halos
    ($10^{10.5} \leq M_{\rm h}/{\rm M_\odot} < 10^{11} \, {\rm M_\odot}$; backsplash excluded) 
    in the DMO simulation TNG300-1-Dark\cite{nelsonIllustrisTNGSimulationsPublic2019}, 
    shown for subsamples
    with different ranges of halo formation time, $z_{\rm f}$ ({\bf a}), 
    and halo spin, $\lambda$ ({\bf b}), and for the total sample 
    (black in {\bf a} and {\bf b}). 
    Fractions of halos in subsamples are 
    equal to those of dwarfs in the subsamples of the massive sample 
    (see Fig.~\ref{fig_b_model}a and Extended Data Table~\ref{tab_gal}).
    {\bf c}, PDF and median 
    of halo concentration ($c$) for halo (sub)samples in {\bf a}.
    Halo concentrations of UDG analogues simulated by 
    NIHAO\cite{DiCintio2017} are shown by 
    grey shaded area (minimum to maximum) and error bar (mean and
    standard deviation).
    Their concentration, $c_{\rm DM}$, is evaluated from the halos in the 
    DMO counterpart of the hydro, compatible with ours.
    {\bf d, e, f}, 2PCCF of central star-forming (${\rm sSFR} \geqslant 10^{-11} {\rm yr}^{-1}$) dwarfs in galaxy-formation models:
    L-Galaxies\cite{henriquesGalaxyFormationPlanck2015}
    (run on TNG100-1-Dark\cite{ayromlouComparingGalaxyFormation2021,nelsonIllustrisTNGSimulationsPublic2019}, {\bf d}), 
    TNG100-1\cite{pillepichFirstResultsIllustrisTNG2018} ({\bf e}) and 
    TNG50-1\cite{pillepichFirstResultsTNG502019} ({\bf f}),
    shown for subsamples with different ranges of $\Sigma_*$, and 
    for the total sample (black). 
    Dwarfs here include those with $10^{8.5} \leqslant M_*/M_\odot < 10^{9}$
    for L-Galaxies and TNG100-1, and $10^{8} \leqslant M_*/M_\odot < 10^{9}$
    for TNG50-1.
    Reference sample includes all galaxies (central or satellite, star-forming 
    or quiescent) above the lower mass limit of the dwarf sample.
    In {\bf a, b, d--f}, grey markers linked by curves from thin to thick
    are the 2PCCFs of massive halos with given ranges of mass in that simulation. 
    Each upper panel shows $w_{\rm p}$, while each lower panel shows the ratio of 
    $w_{\rm p}$ to that of total. 
    Markers with error bars for 2PCCFs show median values with
    $16^{\rm th}$--$84^{\rm th}$ percentiles estimated from 
    bootstrap samples. }
    \label{fig_2pccf_model}
\end{figure*}

\begin{figure}
    \centering
    \includegraphics[scale=0.4]{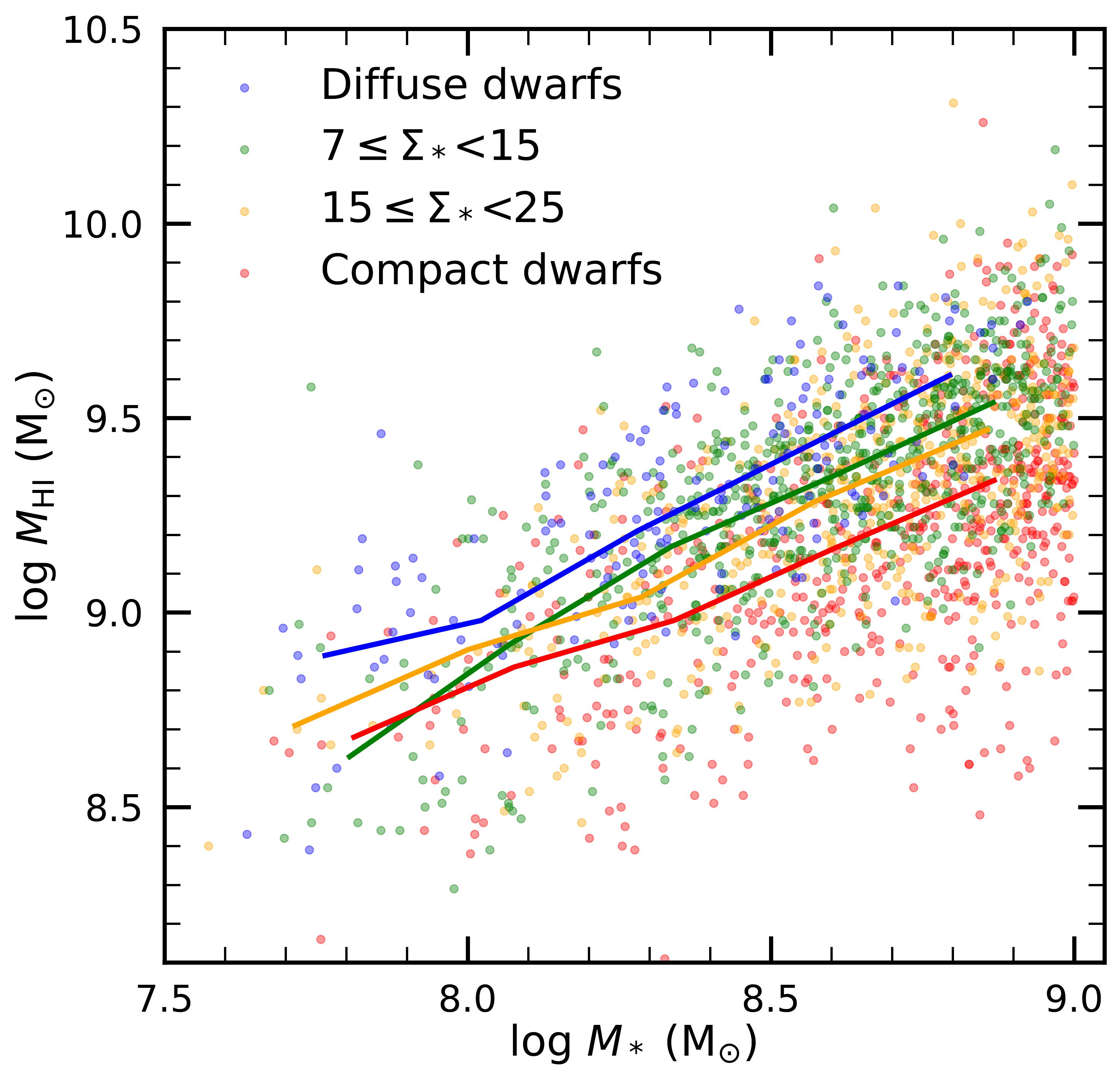}
    \caption{{\bf HI mass ($M_{\rm HI}$) verses $M_*$ for dwarf galaxy samples with varying $\Sigma_*$.} The colored lines represent the median relationships of different samples.}
    \label{fig_HIm}
\end{figure}

\begin{table}
\rmfamily   %% <-- switch to roman font here
\caption{{\bf Sample selection and the corresponding results}}
\label{tab_gal}
\centering
\renewcommand{\arraystretch}{1}
%\resizebox{\textwidth}{!}{
\scalebox{0.7}{
\begin{threeparttable}
\begin{tabular}{c c c c c c c}
\hline
\hline
      Main sample 
    & Sample size
    & $\log\,(M_*/{\rm M_\odot})$\tnote{$\rm a$} 
    & $\log\,(M_{\rm h}/{\rm M_\odot})$\tnote{$\rm b$} 
    & Bias ($z$-matching)\tnote{$\rm d$} 
    & Bias ($z$-weighting)\tnote{$\rm e$} 
    & Halo bias\tnote{$\rm f$}
\\
total         &  $6,919$  & $8.72^{+0.0}_{-0.0}$ &  $10.99^{+0.0}_{-0.0}$ & — & — & $0.68^{+0.0}_{-0.0}$ \\\relax
$0\leq\Sigma_*<7$   &  $349$    & $8.38^{+0.01}_{-0.01}$ &  $10.83^{+0.01}_{-0.01}$ & $2.44_{-0.25}^{+0.25}$ & $2.31_{-0.19}^{+0.20}$ & $0.67^{+0.0}_{-0.0}$ \\\relax
$7\leq\Sigma_*<15$  &  $1,782$  & $8.65^{+0.0}_{-0.0}$ &  $10.96^{+0.0}_{-0.0}$ & $1.41_{-0.12}^{+0.12}$ & $1.49_{-0.11}^{+0.10}$ & $0.68^{+0.0}_{-0.0}$ \\\relax
$15\leq\Sigma_*<25$ &  $1,738$  & $8.73^{+0.0}_{-0.0}$ &  $10.99^{+0.0}_{-0.0}$  & $1.20_{-0.13}^{+0.13}$ & $1.24_{-0.09}^{+0.09}$ & $0.68^{+0.0}_{-0.0}$ \\\relax
$25\leq\Sigma_*$    &  $3,050$  & $8.77^{+0.0}_{-0.0}$ &  $11.01^{+0.0}_{-0.0}$ & $1.00$ & $1.00$ & $0.68^{+0.0}_{-0.0}$ \\
\hline
      Massive sample 
    & Sample size
    & $\log\,(M_*/{\rm M_\odot})$\tnote{$\rm a$} 
    & $\log\,(M_{\rm h}/{\rm M_\odot})$\tnote{$\rm b$} 
    & Bias ($z$-matching)\tnote{$\rm d$} 
    & Bias ($z$-weighting)\tnote{$\rm e$} 
    & Halo bias\tnote{$\rm f$}
\\
total         &  $4,944$  & $8.8^{+0.0}_{-0.0}$ &  $11.03^{+0.0}_{-0.0}$ & — & — & $0.68^{+0.0}_{-0.0}$ \\\relax
$0\leq\Sigma_*<7$   &  $122$    & $8.66^{+0.01}_{-0.01}$ &  $10.96^{+0.01}_{-0.01}$ & $2.22_{-0.33}^{+0.33}$ & $2.4_{-0.28}^{+0.28}$ & $0.68^{+0.0}_{-0.0}$ \\\relax
$7\leq\Sigma_*<15$  &  $1,148$  & $8.76^{+0.0}_{-0.0}$ &  $11.03^{+0.0}_{-0.01}$ & $1.39_{-0.11}^{+0.11}$ & $1.44_{-0.10}^{+0.10}$ & $0.68^{+0.0}_{-0.0}$ \\\relax
$15\leq\Sigma_*<25$ &  $1,284$  & $8.79^{+0.0}_{-0.01}$ &  $11.01^{+0.0}_{-0.0}$  & $1.22_{-0.11}^{+0.11}$ & $1.29_{-0.09}^{+0.09}$ & $0.68^{+0.0}_{-0.0}$ \\\relax
$25\leq\Sigma_*$    &  $2,390$  & $8.82^{+0.0}_{-0.0}$ &  $11.03^{+0.0}_{-0.0}$ & $1.00$ & $1.00$ & $0.68^{+0.0}_{-0.0}$ \\
\hline
      HI-detected\ sample
    & Sample size 
    & $\log\,(M_*/{\rm M_\odot})$\tnote{$\rm a$} 
    & $\log\,(M_{\rm h,HI}/{\rm M_\odot})$\tnote{$\rm c$} 
    & Bias ($z$-matching)\tnote{$\rm d$} 
    & Bias ($z$-weighting)\tnote{$\rm e$} 
    & Halo bias\tnote{$\rm f$}
\\
total         &  $565$  & $8.64^{+0.01}_{-0.01}$ & $10.77^{+0.04}_{-0.04}$
& — & — & $0.69^{+0.01}_{-0.0}$ \\\relax
$0\leq\Sigma_*<7$   &  $59$  & $8.33^{+0.02}_{-0.02}$ & $10.38^{+0.15}_{-0.12}$
& — & — & $0.66^{+0.01}_{-0.01}$ \\\relax
$7\leq\Sigma_*<15$  &  $195$  & $8.55^{+0.01}_{-0.01}$ & $10.72^{+0.06}_{-0.06}$
& — & — & $0.68^{+0.01}_{-0.0}$ \\\relax
$15\leq\Sigma_*<25$ &  $156$  & $8.68^{+0.01}_{-0.01}$ & $10.85^{+0.06}_{-0.07}$
& — & — & $0.69^{+0.01}_{-0.01}$ \\\relax
$25\leq\Sigma_*$    &  $155$  & $8.81^{+0.01}_{-0.01}$ & $10.85^{+0.07}_{-0.08}$
& — & — & $0.7^{+0.01}_{-0.01}$ \\
\hline 
    No $n$-cut\ sample
    & Sample size
    & $\log\,(M_*/{\rm M_\odot})$\tnote{$\rm a$} 
    & $\log\,(M_{\rm h}/{\rm M_\odot})$\tnote{$\rm b$}
    & Bias ($z$-matching)\tnote{$\rm d$} 
    & Bias ($z$-weighting)\tnote{$\rm e$} 
    & Halo bias\tnote{$\rm f$}
\\
total         &  $9,649$  & $8.72^{+0.0}_{-0.0}$ &  $10.98^{+0.0}_{-0.0}$ & — & — & $0.68^{+0.0}_{-0.0}$ \\\relax
$0\leq\Sigma_*<7$   &  $505$    & $8.40^{+0.01}_{-0.01}$ &  $10.83^{+0.01}_{-0.01}$ & $1.77_{-0.15}^{+0.15}$ & $1.83_{-0.13}^{+0.13}$ & $0.67^{+0.0}_{-0.0}$ \\\relax
$7\leq\Sigma_*<15$  &  $2,317$  & $8.67^{+0.0}_{-0.0}$ &  $10.97^{+0.0}_{-0.0}$ & $1.32_{-0.09}^{+0.09}$ & $1.27_{-0.07}^{+0.07}$ & $0.68^{+0.0}_{-0.0}$ \\\relax
$15\leq\Sigma_*<25$ &  $2,122$  & $8.74^{+0.0}_{-0.0}$ &  $11.00^{+0.0}_{-0.0}$  & $1.10_{-0.09}^{+0.09}$ & $1.12_{-0.07}^{+0.07}$ & $0.68^{+0.0}_{-0.0}$ \\\relax
$25\leq\Sigma_*$    &  $4,705$  & $8.76^{+0.0}_{-0.0}$ &  $11.00^{+0.0}_{-0.0}$ & $1.00$ & $1.00$ & $0.68^{+0.0}_{-0.0}$ \\

\hline
\hline
\end{tabular}
\caption{{\bf Sample selection and the corresponding results} The columns show the values of the corresponding quantities, with uncertainties corresponding to the $16\%$ and $84\%$ percentiles. The uncertainties are rounded to two decimal places, and a value of $0.0$ represents the uncertainty of less than $0.004$. Column a: Median stellar mass of the sample; Column b: Halo mass estimated from SHMR; Column c: Halo mass measured from\ HI-kinematics; Column d: Relative bias obtained using the $z$-matching method; Column e: Relative bias obtained using the $z$-weighting method; Column f: Theoretical halo bias of the sample.
}

\end{threeparttable}}
\end{table}

\clearpage

\newenvironment{supplementary}{%
\section*{Supplementary Information}%
\setlength{\parskip}{12pt}%
}{}

\begin{supplementary}\label{sec_supplementary}

\renewcommand{\figurename}{\bf Supplementary Fig.}
\setcounter{figure}{0}
\renewcommand{\tablename}{\bf Supplementary Table}
\setcounter{table}{0}

\subsection{The details of the sample selection and splitting methods}\label{sec_sample_detail}

We opted for this specific method of sample splitting because the diffuse dwarfs defined in this paper are akin to ultra-diffuse galaxies (UDGs) in the literature\cite{vanDokkum2015, Koda2015, DiCintio2017, RongY2017}. 
UDGs are identified with thresholds of a surface brightness of $\mu_{\rm e}> 24\, {\rm mag/arcsec^{2}}$ 
and an effective radius $R_{50}>1.5\, {\rm {kpc}}$\cite{vanDokkum2015}. Using the relation between stellar mass-to-light 
ratio and color (MLCR)\cite{Bell2003}, we obtain
\begin{align}
    \log\left( M_*/{\rm M_\odot} \right) &=-0.306+1.097(g-r)-0.1-0.4(M_{r}-4.64)-0.12 \nonumber \\
                    &= 1.33+1.097(g-r)-0.4M_r \,,
\end{align}
where $M_r$ is the $r$-band absolute magnitude, 4.64 is the $r$-band magnitude of the Sun in the 
AB system\cite{Blanton2007}, the $-0.10$ term effectively implies the use of a Kroupa (2001) IMF\cite{Kroupa2001}, which is adopted for the estimation of $M_*$\cite{Kauffmann2003}, 
and the $-0.12$ term is used to account for the difference between the MPA-JHU mass and the mass 
estimated using the MLCR at $\log (M_*/{\rm M_\odot})\sim9.0$ (see Figure 8 in Zhang et al.\cite{Zhang2022}). 
Please see Yang et al.\cite{Yang2007} for details. We then obtain,
\begin{equation}
    \log\frac{\Sigma_*}{{\rm M_\odot}{\rm pc}^{-2}}=9.96+1.097(g-r)+4\log(1+z)-0.4\frac{\mu_{\rm e}}{\rm mag/arcsec^{2}}\\\,.\label{eq_Sigma}
\end{equation}
Assuming $g-r=0.6$ (UDGs in clusters are usually red), the surface brightness criterion of 
$\mu_{\rm e}=24~\rm mag/arcsec^{2}$ roughly corresponds to $\Sigma_*=10\,{\rm M_\odot}{\rm pc}^{-2}$. 
Assuming $g-r=0.3$ (UDGs in fields are usually blue), the surface brightness criterion of 
$\mu_{\rm e}=24~\rm mag/arcsec^{2}$ roughly corresponds to $\Sigma_*=5\,{\rm M_\odot}{\rm pc}^{-2}$. 
We therefore adopted $\Sigma_*<7\,{\rm M_\odot}{\rm pc}^{-2}$ to select diffuse dwarfs. We also checked the 
$R_{50}$-distribution of diffuse dwarfs and found that $321/349$ exhibit $R_{50}>1.5\, {\rm {kpc}}$.
Furthermore, the S\'ersic index distribution of UDGs usually peaks at $n\approx1$\cite{Koda2015,Greco2018}, 
indicating an exponential light profile. Our criterion of $n<1.6$ results in a median S\'ersic index 
of $\sim1.2$ for these dwarfs, akin to UDGs, and ensures a relatively pure sample of late-type 
morphology. High-resolution images\cite{aiharaHyperSuprimeCamSSP2018,miyazakiHyperSuprimeCamSystem2018} were cross-matched with our sample and visually inspected to verify the purity of the selection.

\subsection{Control-sample construction in the $z$-matching method}\label{sec_control_sample}

In the $z$-matching scheme, we constructed a control sample for each sample in such a 
way that all control samples share the same redshift distribution, $f_{\rm ctl}(z)$.
To do this, we first determined $f_{\rm ctl}(z)$ through 
\begin{equation}
    f_{\rm ctl}(z) = {\rm min}(f_{1}(z), f_{2}(z),...,f_{n}(z)),
\end{equation}
where $f_{x}(z)$, with $x=1,2,...,n$, is the redshift distribution of the 
$x$th sample. The shaded region in Extended Data Fig.~\ref{fig_para_dis}d shows $f_{\rm ctl}(z)$.
We then computed the numbers for the control sample $x$ within a redshift bin $z$ using
\begin{equation}
    n_{x,\rm ctl}(z)=f_{\rm ctl}(z)/f_{x}(z)*n_x(z),
\end{equation}
where $n_x(z)$ is the number of galaxies in the $x$th original sample within the same 
redshift bin. Since $f_{\rm ctl}(z)\leq f_x(z)$, one has $n_{x,\rm ctl}(z)\leq n_x(z)$.
Finally, we randomly chose $n_{x,\rm ctl}(z)$ galaxies from the original sample in the corresponding 
redshift bin to create the control sample. 

\subsection{The impact of uncertainties in $M_*$, $R_{50}$ and $z$}\label{sec_uncert} 

In the main text, stellar masses from the MPA-JHU catalog are adopted, and  
the typical statistical uncertainty in the mass is about $0.08\, {\rm dex}$.
The GSWLC catalog\cite{Salim2016} also provides stellar-mass estimates (hereafter the GSWLC mass) 
based on the UV$+$optical$+$mid-IR SED fitting. The two masses are tightly correlated, 
but there is a systematic offset of $0.17\, {\rm dex}$ which increases with increasing stellar mass. 
The scatter of the relation and the offset between the two mass estimates are larger than the 
statistical uncertainties in the MPA-JHU masses, signifying a potential 
issue in our analysis. 

\begin{table*}
    \centering
    \caption{Sample selection and the corresponding results based on GSWLC stellar mass}
    \scalebox{0.7}{
    \begin{threeparttable}
    \begin{tabular}{c c c c c c c}
    \hline
    \hline
    $\rm GSWLC\ samples$ 
    & $\rm Sample\ size$ 
    & $\log$$(M_*/{\rm M_\odot})$\tnote{$\rm a$} 
    & $\log$$(M_{\rm h}/{\rm M_\odot})$\tnote{$\rm b$} 
    & $\rm Bias$ $z$-$\rm matching$\tnote{$\rm c$} 
    & $\rm Bias$ $z$-$\rm weighting$\tnote{$\rm d$} 
    & $\rm Halo\ bias$\tnote{$\rm e$}\\
    $\rm total$         &  $4,699$  & $8.75_{-0.0}^{+0.0}$ & $11.0_{-0.05}^{+0.05}$ & - & - & $0.68_{-0.0}^{+0.0}$ \\\relax
    $0\leq\Sigma_*<10$   &  $262$   & $8.49_{-0.01}^{+0.01}$ & $10.89_{-0.08}^{+0.06}$ & $2.32_{-0.36}^{+0.36}$ & $1.95_{-0.23}^{+0.22}$ & $0.67_{-0.0}^{+0.0}$ \\\relax
    $10\leq\Sigma_*<20$  &  $1,109$  & $8.71_{-0.0}^{+0.0}$ & $10.99_{-0.07}^{+0.05}$ & $1.93_{-0.21}^{+0.21}$ & $1.75_{-0.13}^{+0.13}$ & $0.68_{-0.0}^{+0.0}$ \\\relax
    $20\leq\Sigma_*<40$ &  $1,542$  & $8.77_{-0.0}^{+0.0}$ & $11.0_{-0.06}^{+0.06}$ & $1.2_{-0.15}^{+0.15}$ & $1.15_{-0.1}^{+0.09}$ & $0.68_{-0.0}^{+0.0}$ \\\relax
    $40\leq\Sigma_*$    &  $1,786$  & $8.78_{-0.0}^{+0.0}$ & $11.01_{-0.05}^{+0.06}$ & $1.0_{-0.0}^{+0.0}$ & $1.0_{-0.0}^{+0.0}$ & $0.68_{-0.0}^{+0.0}$ \\
    \hline
    \hline
    \end{tabular}
    \caption*{The columns show the values of the corresponding quantities, with uncertainties corresponding to the 16\% and 84\% percentiles. The uncertainties are rounded to two decimal places, and a value of 0.0 represents the uncertainty of less than 0.004. Column a: Median stellar mass of the sample; Column b: Halo mass estimated from SHMR; Column c: Relative bias obtained using the $z$-matching method; Column d: Relative bias obtained using the $z$-weighting method; Column e: Theoretical halo bias of the sample.}
    \end{threeparttable}}
    \label{tab_1}
\end{table*}

%\begin{figure}
%    \centering
%    \includegraphics[scale=0.2]{Figs/fig_bias3.pdf}
%    \caption{Relative bias as a function of $\Sigma_*$ for dwarf galaxies. Here we adopt the GSWLC stellar mass\cite{Salim2016} instead of the MPA-JHU mass. The $^{0.1}(g-r)$ color and the S\'ersic index cuts were used to select the dwarf samples. Please see the text for the details of sample selection.  
%    The relative bias and its corresponding lower and upper error bars correspond to the median and the $16^{\rm th}$ and $84^{\rm th}$ percentiles of the posterior distribution obtained by MCMC fitting, respectively. }
%    \label{fig_gswlcbias}
%\end{figure}

To address this issue we constructed a new dwarf sample based on the GSWLC mass estimates
using $7.5\leq\log M_*/{\rm M_\odot} < 9$, $^{0.1}(g-r) < 0.6$ and $n<2.0$. 
The total dwarf sample is divided into four subsamples with $0 \leq \Sigma_* < 10$ (diffuse), 
$10 \leq \Sigma_* < 20$, $20 \leq \Sigma_* < 40$, and $40 \leq \Sigma_*$ (compact), respectively. 
Since the GSWLC mass is greater than the MPA-JHU mass by $0.17 \, {\rm dex}$ at $\log M_*/{\rm M_\odot}\sim9$, 
we adopted a higher threshold, $7\,{\rm M_\odot}{\rm pc}^{-2}\times 10^{0.17}\simeq 10\,{\rm M_\odot}{\rm pc}^{-2}$, to select diffuse dwarfs.
This gives $262$ diffuse dwarfs and $1,786$ compact dwarfs. 
The relative bias obtained from these subsamples is listed in Supplementary
Table~\ref{tab_1}. The diffuse dwarfs still have significantly 
higher bias than compact dwarfs. The relative bias for the diffuse sample is around $2$, 
similar to the result based on the MPA-JHU mass but with larger errors.   
The reason for this is that a significant fraction of dwarfs do not have 
GSWLC mass estimates; the fraction is as high as $26\%$ for diffuse dwarfs defined by the MPA-JHU mass.
We thus conclude that our results are not sensitive to the stellar mass measurements.

The NYU-VAGC catalog does not provide uncertainties for the $R_{50}$ measurements. 
The values of $R_{50}$ given by the catalog are derived from the S\'ersic-model fitting\cite{Blanton2005b}
and their uncertainties, shown in Figure 10 of the cited reference, are very small, typically about $10\%$.
We thus believe that size uncertainties do not affect our results significantly.
The uncertainties in redshift are very small, typically with $\Delta z/(1+z)=2\times10^{-5}$, 
corresponding to a negligible distance uncertainty of $\Delta r=c\Delta z/H_0=0.06\, h^{-1}{\rm {Mpc}}$. 
The typical redshift of our samples galaxies is $z\sim 0.02$, corresponding to receding 
velocity of $\sim 6,000\,{\rm km\,s^{-1}}$. Thus, peculiar velocities of galaxies 
may have a sizable effect on the estimates of their stellar masses.  
The impact of the uncertainties in the stellar mass estimates has been tested 
above.

\subsection{The distances to nearest groups} \label{sec_distance_to_group}

Backsplash halos, which 
were once contained in massive halos but now are independent, 
make significant contributions to the halo assembly bias\cite{Wang2009}. 
Most of the backsplash halos reside within three to four times the virial radius of their host 
halos\cite{Wang2009}. The distance distribution of backsplash halos from their host halos 
reaches its maximum at less than twice the virial radius of their hosts\cite{Wang2009}. 
Moreover, since the projected distance is smaller than the 3-D distance, 
the median projected distance of diffuse dwarfs to nearby groups should be smaller than two 
times the virial radius if the dwarf sample is dominated by backsplash halos.
To test this, we identified, for each diffuse dwarf galaxy,  the neighboring groups with 
$|\Delta v| \leq 3 v_{\rm vir}$, where $\Delta v$ is the line-of-sight velocity difference 
between the dwarf and the group, and $v_{\rm vir}$ is the virial velocity of the group. 
We computed the projected separation ($R_{\rm sep}$) between the dwarf and neighboring groups 
and select the nearest group as the one with the smallest $R_{\rm sep}/R_{\rm vir}$, 
where $R_{\rm vir}$ is the virial radius of the group. 
We found that the median $R_{\rm sep}$ to the nearest groups with $M_{\rm h}>10^{12}\, {\rm M_\odot}$ 
is $1.84\, h^{-1}{\rm {Mpc}}$, about $4.8$ times the virial radius, while the median $R_{\rm sep}$ from the 
nearest groups with $M_{\rm h}>10^{13}\, {\rm M_\odot}$ is about $5.9\, h^{-1}{\rm {Mpc}}$, about $8.0$ times the virial radius. 
These large separations are in conflict with associating diffuse dwarfs with backsplash halos.

\subsection{Details of the halo-mass estimate from HI kinematics} \label{sec_detail_of_HI_mh}

We used the same method as described in Guo et al. (2020)\cite{Guo2020-UDG} to evaluate 
the $20\%$ peak width of the HI line width ($W_{20}$) from the HI spectrum for each galaxy. Since resolved HI maps 
are not available, we assumed the inclination of the HI disk, $\phi$, to be the same as that of the stellar 
disk, ${\sin \phi} = \sqrt{[1-(b/a)^2]/(1-q_0^2)}$, where $q_0\sim 0.2$\cite{Rong24}. 
The circular velocity $V_{\rm{c}}$ is then estimated as $V_{\rm{c}}=W_{20}/(2\sin\phi)$. For a typical dwarf galaxy, the circular velocity at a large radius, such as the HI radius $r_{\rm{HI}}$ 
(defined as the radius at which the HI surface density attains $1\ \rm M_{\odot}\rm{pc^{-2}}$), 
is expected to be $V_{\rm{c}}$. The dynamical mass enclosed within $r_{\rm{HI}}$ is
\begin{equation}
 M_{\rm dyn} (<r_{\rm HI}) =V_{\rm c}^2 r_{\rm HI}/G\,,
 \label{dm}
\end{equation}
where $G$ is the gravitational constant. The estimation of $r_{\rm{HI}}$ is facilitated by the tight 
correlation between $r_{\rm{HI}}$ and HI mass $M_{\rm{HI}}$ inferred from 
observations: $\log_{10} r_{\rm{HI}}=0.51\log_{10} M_{\rm{HI}}-3.59$\cite{Wang16,Gault21}.

Assuming a Burkert profile\cite{Burkert95} with a central core\cite{Marchesini02,Rong24a}, we can estimate 
the halo mass using 
 \begin{equation}
\begin{aligned}
{M_{\rm dyn}(<r_{\rm HI})}-M_{\rm{bar}}& =  \int_0^{r_{\rm{HI}}}4\pi r^2\rho_{\rm{B}}(r)\mathrm{d}r \\
 & =  2\pi \rho'_{0} r_{0}^3\left[\ln \left(1+\frac{r_{\rm{HI}}}{r_{0}}\right)
 +0.5\ln\left(1+\frac{r_{\rm{HI}}^2}{r_{0}^2}\right) 
 -{\rm arctan}\left(\frac{r_{\rm{HI}}}{r_{0}}\right)\right]\,,
\end{aligned}
  \label{bur}
  \end{equation}
and
\begin{equation}
	\begin{aligned}
M_{200\rm{c}} & = \int_0^{R_{200\rm c}}4\pi r^2\rho_{\rm B}(r)\mathrm{d}r \\
& =  2\pi \rho'_{0} r_{0}^3\left[\ln \left(1+\frac{R_{200\rm c}}{r_{0}}\right)+0.5\ln\left(1+\frac{R_{200\rm c}^2}{r_{0}^2}\right) 
-{\rm arctan}\left(\frac{R_{200\rm{c}}}{r_{0}}\right)\right]\,,
\end{aligned}
\label{bur_m}
\end{equation}
where $M_{\rm{bar}}\simeq M_{*}+1.33M_{\rm{HI}}$ denotes the galactic baryonic mass; 
$r_{0}$ and $\rho'_{0}$ are free parameters describing the core of the dark matter halo, 
and $r_0$ is found to be related to $M_{200\rm{c}}$ by \cite{Salucci07},
\begin{equation}
\log [(r_0/{\rm kpc})] = 0.66-0.58(\log[M_{200\rm{c}}/10^{11} \rm M_{\odot}])\,.
\label{bur_c}
\end{equation}
The halo mass, $M_{\rm 200c}$, can then be estimated with equations~(\ref{bur}), (\ref{bur_m}), and (\ref{bur_c}). 
Note that $R_{\rm 200c}$ represents the virial radius enclosing a mean density that is 
200 times the critical value and $M_{\rm 200c}$ is the mass within $R_{\rm 200c}$.

The uncertainty of the halo mass is determined using a Monte Carlo method, taking into account 
uncertainties in the baryonic mass ($\sigma_{M_{\rm{bar}}}$), in $r_{\rm{HI}}$ ($\sigma_{r_{\rm{HI}}}$), 
and in $W_{20}$ ($\sigma_{W_{20}}$), as well as potential misalignment between the HI and 
stellar inclinations, and the uncertainty in the $r_0$-$M_{200\rm{c}}$ relation. 
The error term $\sigma_{M_{\rm{bar}}}$ also includes uncertainties in the HI mass, 
$\sigma_{M_{\rm{HI}}}$, 
provided by ALFALFA\cite{Haynes2018}, and in the stellar mass $\sigma_{M_{*}}$ due to 
uncertainties in the distance and magnitude. Therefore, $\sigma^2_{M_{\rm{bar}}}=\sigma^2_{M_{*}}+(1.33\sigma_{M_{\rm{HI}}})^2$. 
The uncertainty $\sigma_{r_{\rm{HI}}}$ is estimated based on the HI mass error, while $\sigma_{W_{20}}$  
follows the method outlined in Guo et al.\cite{Guo2020-UDG}. Previous studies have shown that the stellar and gas disks in galaxies may not be perfectly co-planar, often exhibiting a small 
inclination difference of ${\rm \delta}\phi < 20^\circ$\cite{Starkenburg-counterrotation, Guo2020-UDG, Gault2021-VLA_UDG}. 
To address this misalignment, we assumed that ${\rm \delta}\phi$ follows a Gaussian distribution 
centered at $0^\circ$ with a standard deviation of $\sigma_{\phi} = 20^\circ$ to represent 
the uncertainty associated with $\phi$. 
For each galaxy, we generated $1,000$ sets of ($M_{\rm{bar}}$, $W_{20}$, $\phi$, and $r_{\rm{HI}}$) based on their 
average values and associated uncertainties. We assumed Gaussian distributions for these parameters 
centered at their average values, with the 1-$\sigma$ ranges matching the uncertainties. 
Consequently, we obtained 1,000 halo masses using equations~(\ref{dm})--(\ref{bur_c}). The standard 
deviation of these halo masses ($\sigma_{M_{200}}$) is combined with the uncertainty of the 
$r_0$-$M_{200\rm{c}}$ relation to determine the final halo mass uncertainty, as
\begin{equation}
    \sigma'_{M_{\rm 200c}} = \sqrt{\sigma_{M_{\rm 200c}}^2 + (0.15~{\rm dex})^2}\,, 
\end{equation}
where the scatter of the dark mass profile for a given halo mass\cite{Wangj20} 
is approximately $0.15\, {\rm dex}$, which is used for the uncertainty in the $r_0$-$M_{200\rm{c}}$ relation.

To compare with the mass derived from SHMR, we coverted $M_{\rm{200c}}$ into $M_{\rm 200m}$ 
by using the derived Burkert profile.

\subsection{Abundance matching} \label{sec_AM}

In Methods, we showed the $\Sigma_*$-$z_{\rm f}$ mapping based on abundance matching. Here we provide further details.

For abundance matching to work correctly, the procedure must 
(i) preserve the rank order of $z_{\rm f}$ and $\Sigma_*$ to the degree 
set by the adopted scatter and (ii) yield a $\Sigma_*$-distribution for halos 
consistent with that of dwarf galaxies. 
Our mapping formula meets these criteria.

The mapping first applies $\mathcal{P}_{z_{\rm f}}$, which transforms $z_{\rm f}$ 
into a uniformly distributed variable over $[0, 1]$. Next, $\mathcal{N}^{-1}$ 
converts this uniform variable into a unit Gaussian variable. 
The composition $\mathcal{N}^{-1}\circ \mathcal{P}_{z_{\rm f}}$ 
thus maps the $z_{\rm f}$-distribution into a unit Gaussian distribution. 
A unit Gaussian random scatter, $\epsilon$, is then added, 
where a correlation coefficient $\rho$ controls the weight. 
By the additive property of Gaussian variables, the result remains a unit Gaussian variable 
and has correlation coefficients $\rho$ and $\sqrt{1-\rho^2}$ 
with the original Gaussian variable and $\epsilon$, respectively. 
Finally, $\mathcal{N}$ transforms the unit Gaussian back to a uniform variable, 
which is then converted to $\Sigma_*$  that follows the distribution of dwarfs. 
Here, $\rho$ controls the scatter: $\rho=1$ indicates perfect correlation between
$z_{\rm f}$ and $\Sigma_*$, whereas $\rho=0$ implies no correlation.

This mapping is mathematically concise. However, it (i) does not allow the 
scatter to vary with  $z_{\rm f}$ and (ii) does not quantify the scatter 
as intuitively as directly adding scatter to a physical variable such as $\Sigma_*$. 
To address (i), Fig.~\ref{fig_b_model}a shows results for a number of $\rho$ values. 
For $\Sigma_* \gtrsim 10\,{\rm M}_\odot{\rm pc}^{-2}$, $\rho \gtrsim 0.5$ yields a match to observations, while for $\Sigma_* \lesssim 10\,{\rm M}_\odot{\rm pc}^{-2}$ (``diffuse dwarfs''), a stronger correlation ($\rho \gtrsim 0.8$) is required. 
To address (ii), Fig.~\ref{fig_b_model}b displays the median and $16^{\rm th}$--$84^{\rm th}$ percentile range of the $z_{\rm f}$-$\Sigma_*$ relation for $\rho = 0.85$
(the case that appears closely matching to the observed bias-$\Sigma_*$ relation). 
The percentile range intuitively quantifies the introduced scatter by the abundance matching.

\subsection{Model assumptions, results and implications} \label{sec_model_assumptions}

The figures in the main texts (Figs.~\ref{fig_bias}--\ref{fig_sidm_bias}) were arranged in a logical order,
each building upon the previous one with progressively more assumptions
and leading to increasingly refined results and implications. 
The conclusions can thus be judged incrementally, based on the robustness of the
assumptions introduced at each step. Below is a brief summary.

\noindent
(i) In Fig.~\ref{fig_bias}, we presented observational results. Here, assumptions include 
sample selection and property measurements.  
The results show that diffuse dwarfs have stronger clustering than other 
isolated dwarfs.

\noindent
(ii) In Fig.~\ref{fig_cosmicweb}, we reconstructed the density field at $z\approx 0$ using 
the SDSS sample. Assumptions include the group finder, RSD correction, and 
halo-matter cross-correlation. 
The results show a strong correlation of diffuse dwarfs with filaments/knots, 
and an anti-correction with voids. 
This implies that diffuse dwarfs preferentially reside within/around 
large cosmic structures, constraining the conditions for their formation.

\noindent
(iii) In Fig.~\ref{fig_b_model}, we traced the evolution of the $z\approx 0$ density field 
back to high $z$ using a constrained simulation, ELUCID. As the simulation 
cannot resolve assembly of individual halo, we introduced an abundance modeling 
between $z_{\rm f}$ and $\Sigma_*$. The assumptions here are the 
initial-condition reconstruction and the abundance modeling. 
The results show that the $z_{\rm f}$-bias of halos can explain the 
observed $\Sigma_*$-bias of dwarfs, raising the question of how $\Sigma_*$ 
is physically linked to $z_{\rm f}$. This result also provides clues 
for revisions to existing models in $\Lambda$CDM cosmology. 
The data product of this step is the $\Sigma_*$ assigned to each dwarf-host 
halo within the ELUCID volume.

\noindent
(iv) In Fig.~\ref{fig_sidm_bias}, we introduced SIDM as a possible explanation for our findings. 
The assumption in this step is the isothermal Jeans model. 
Each halo, with its $\Sigma_*$ assigned as above, is thus predicted to contain 
a SIDM core characterized by $r_{\rm c}$, $r_{\rho_0/4}$, and $\rho_0$. 
The results show a similarity between SIDM cores and dwarfs, 
providing insights for future observations and theoretical studies. 
Parameterizing $R_{50} = A_{\rm r}r_{\rm c}$, the predicted 
bias-$\Sigma_*$ relations by our SIDM model are shown in Fig.~\ref{fig_b_model} for 
comparison with the observation and other models. 

\end{supplementary}

%% Put the bibliography here, most people will use BiBTeX in
%% which case the environment below should be replaced with
%% the \bibliography{} command.

\newcommand\urlprefix{}
\bibliographystyle{naturemag}
\bibliography{main.bib}

\end{document}